\documentclass[11pt,notitlepage]{article}
\usepackage[utf8]{inputenc}
\usepackage{amsmath,amsthm,amsfonts,amssymb,amscd}
\usepackage{kotex}
\usepackage[colorlinks=false]{hyperref}
\usepackage{multirow,booktabs}
\usepackage{fancyvrb} % code input
\usepackage[table]{xcolor}
\usepackage{fullpage}
\usepackage{subfig, bm}
\usepackage{mathrsfs}
\usepackage[font=footnotesize,labelfont=bf]{caption}
\usepackage{import}
\usepackage{lastpage, mathtools}
\usepackage{changepage}% http://ctan.org/pkg/changepage
\usepackage{dsfont}
\usepackage{cancel}
\usepackage{url}
\usepackage{mdwlist}
\usepackage{enumitem}
\usepackage{fancyhdr}
\usepackage{sectsty}
\usepackage{mathrsfs}
\usepackage{algorithm}
\usepackage{algorithmic}
\usepackage{wrapfig}
\usepackage[numbers]{natbib}
\usepackage{setspace}
\usepackage{calc}
\usepackage{multicol}
\usepackage{cancel}
\usepackage[retainorgcmds]{IEEEtrantools}
\usepackage{xfrac}
\usepackage[margin=3cm]{geometry}
\usepackage{amsmath}

\usepackage{empheq}
\usepackage{framed}
\usepackage{etoolbox}
\usepackage{float}
\usepackage{tikz}
\usetikzlibrary{matrix,decorations.pathreplacing}
\usepackage[most]{tcolorbox}
\usepackage{xcolor}
\colorlet{shadecolor}{blue!15}
\usepackage{subfiles}
\usepackage{sjmacros}
\usepackage{thmstyles}
\allowdisplaybreaks

\title{\LARGE \bf A Two-Part Controller Synthesis Approach for Nonlinear Stochastic Systems Perturbed by L\'{e}vy Noise\\ Using Renewal Theory and HJB-Based Impulse Control}

\author{
SooJean Han\thanks{SooJean Han and John C. Doyle are with the Department of Computing and Mathematical Sciences, California Institute of Technology, Pasadena, CA 91125, USA. Emails: {\tt\small soojean@caltech.edu}, {\tt\small doyle@caltech.edu}.}
\and Soon-Jo Chung\thanks{Soon-Jo Chung is with the Department of Computing and Mathematical Sciences and Department of Aerospace, California Institute of Technology, Pasadena, CA 91125, USA. Email: {\tt\small sjchung@caltech.edu}.}
}
\date{}

\begin{document}
\maketitle
\setcounter{section}{0}
\thispagestyle{empty}

\begin{abstract}
    We are motivated by the lack of discussion surrounding methodological control design procedures for nonlinear shot and Lévy noise stochastic systems to propose a hierarchical controller synthesis method with two parts. The first part is a primitive pattern-learning component which recognizes specific state sequences and stores in memory the corresponding control action that needs to be taken when the sequence has occurred. The second part is a modulation control component which computes the optimal control action for a pattern when it has occurred for the first time. Throughout our presentation of both components, we provide a self-contained discussion of theoretical concepts from Poisson processes theory, renewal theory, and impulse control, all of which are necessary as background. We demonstrate application of this controller to the simplified, concrete case studies of fault-tolerance and vehicle congestion control.
\end{abstract}

% \violet{Other citations:\cite{hildebrandt59}.}

%%%%%%%%%%%%%%%%%%%%%%%%%%%%%%%%%%%%%%%%%%%%%%%%%%%%%%%%%%%%%%%%%%%%%%%%%%%%%%%%%%%%%%%%%%%%%%%%%%%%%%
%%%%%%%%%%%%%%%%%%%%%%%%%%%%%%%%%%%%%%%%%%%%%%%%%%%%%%%%%%%%%%%%%%%%%%%%%%%%%%%%%%%%%%%%%%%%%%%%%%%%%%

\section{Introduction}\label{sec:intro}
% \blue{
% Things to fix:
% \begin{enumerate}
%     % \item Introduction reads too arrogantly. Write less like novel.
%     % \item \textbf{[Added 6/28]} Conclusion needs mention of future work and limitations.
%     % \item More discussion/relationship on the background of HJB optimal control before impulse control section.
%     % \item \textbf{[Added 6/28]} For both the 1D reference-tracking example and the vehicle traffic control example, only the first part learning component of the controller was considered. Impulse control wasn’t mentioned.
%     % \item Change numbers in Ross’ examples. Refer to Ross for proof of Ignatov’s theorem. Cite Ross' book and my own book more scattered throughout.
%     % \item Link the vehicle traffic control example with the application of internet.
%     % \item Emphasize that the novel contribution of the paper is the two-part construction of the controller, the learning component in order to reduce redundancy in computation. Impulse control $\to$ moderation control.
%     % \item Remark about practicality aspect of impulse control: how impulse isn’t actually impulse, but hill. Closest practical working example is bang-bang control in spacecraft fault tolerance.
% \end{enumerate}
% }

The study of \textit{dynamical systems} in the field of control theory utilizes mathematical models to describe the evolution of a certain set of system quantities over time. In its most abstract form, a dynamical system is traditionally written as the following ODE:
\begin{align}\label{eq:basic_ode}
	d\xvect(t) = f(t,\xvect)dt
\end{align}
where $f: \Rbb^{+}\times\Rbb^n \to \Rbb^n$ is a deterministic function in $\Ccal^{(1,2)}$, i.e., $f$ is continuously-differentiable in time and twice continuously-differentiable in state, and the state dimension is $n\in\Zbb^{+}$. Core theoretical foundations for dynamical system models have been studied for decades and gathered in numerous comprehensive references such as~\cite{khalil_book,sastry_book,slotine_li_book,zhou_doyle_book} over time. 
% These foundations provide invaluable mathematical insights towards solving the three most common problems which arise for any type of real-world dynamical system: 1) determining the stability of the system, 2) controlling the system (controller design), and 3) observing the state of the system (observer design). The observer design problem is often considered to be the dual problem to the control design problem.

While much of basic control theory is studied under the condition that $f$ in~\eqn{basic_ode} is a deterministic function of $t$ and $\xvect$, many dynamical systems in practice are also affected by some amount of randomness, which prevents future states of the system from being able to be predicted precisely. These \textit{stochastic (dynamical) systems} come in many different types and arise in a wide number of disciplines; some examples include voltage fluctuations in metallic conductors~\cite{beenakker92} in fields of physics, photon-counting in the field of optics~\cite{ligo11}, and the regulation of control circuits in biomolecular systems~\cite{olsman19}. Motivated by the existence of these systems, there has been a vast literature of references, such as~\cite{oksendal_book,pinsky11book,karlin76book,karlin81book}, dedicated to the rigorous study of stochastic systems and random processes. 
% The three common problems for dynamical systems mentioned previously are also valid to ask for stochastic systems, and are typically referred to as the problems of 1) \textit{stochastic stability}, 2) \textit{stochastic control}, and 3) \textit{stochastic (state) estimation} in control theory literature.

% When addressing the problems of stochastic systems, 
When considering the problem of control or state-estimation for stochastic systems, many designs for controllers or observers, especially model-based designs, typically aim for robustness against specifically additive Gaussian white noise (AWGN). The model of such a stochastic system is expressed as the following SDE:
\begin{align}\label{eq:gen_white_sde}
    d\xvect(t) = f(t, \xvect)dt + \sigma(t, \xvect)dW(t)
\end{align}
where 
\begin{itemize}
    \item $W: \Rbb^{+}\to\Rbb^d$ is a $d$-dimensional standard Brownian motion process.

    \item $\sigma: \Rbb^{+}\times\Rbb^n \to \Rbb^{n\times d}$, $\sigma \in \Ccal^{(1,2)}$ is the variation of the Gaussian white noise.
\end{itemize}
There are many applications such as vision-based localization/mapping~\cite{yang13}, spacecraft navigation~\cite{capuano19}, and motion-planning~\cite{kalakrishnan11} where using Gaussian white noise processes in the model is justifiable in practice. From a theoretical perspective, Gaussian white noise processes are appealing to study because their convenient properties (e.g., continuity of sample paths, normality of the distribution, the Central Limit Theorem) make the stochastic system easier to analyze than non-Gaussian stochastic systems. Consequently, there is a wealth of literature that has been devoted towards analyzing the stability of systems with AWGN perturbations, and designing controllers and observers for them. Some classical model-based methods are the Linear Quadratic Gaussian (LQG) model~\cite{bernstein88,doyle78}, as well as Kalman filtering~\cite{kalman60} and its extensions~\cite{reif99,wan00}. More recent methods of model-based controller and observer designs for Gaussian white noise include the path integral approach~\cite{theodorou10}, convex optimization-based approaches~\cite{rawlik13,nakka19}, as well as a number of reinforcement learning based approaches~\cite{munos98,deisenroth15}.

However, there is a major lack of generality with the assumption of Gaussian white noise. Because it is small in magnitude and continuous in the sense that changes occur gradually over a measurable duration of time, it is unable to account for sudden impulsive perturbations. One particular class of non-Gaussian noise which is suitable for modeling impulsive perturbations is \textit{Poisson shot noise}~\cite{baccelli_book}. The dynamics of a stochastic system where shot noise is injected additively is written as the following SDE:
\begin{align}\label{eq:gen_shot_sde}
    d\xvect(t) = f(t, \xvect)dt + \int_{\Rbb^{\ell}}\xi(t, \xvect, \zvect)N(dt,d\zvect)
\end{align}
where
\begin{itemize}
    \item $N(dt,d\zvect)$ is a $r$-dimensional standard Poisson process with some intensity $\lambda\in\Rbb^{r}$.

    \item $\xi: \Rbb^{+}\times\Rbb^n\times\Rbb^{\ell} \to \Rbb^{n\times r}$, $\xi\in\Ccal^{(1,2,1)}$ is a function which assigns weights to the individual components of $N$, since the standard Poisson process itself has unit weights. Essentially, $\xi$ is a function that describes the impulsive jumps that occurs in the system.
\end{itemize}

Poisson shot noise arises in a wide diversity of real-world applications almost just as frequently as Gaussian white noise does. Some examples include the high fluctuations in stock prices within the field of finance~\cite{cont06}, as well as the neuronal spikes that come from monitoring brain activity within the field of neuroscience~\cite{patel09} are better modeled as impulsive perturbations than as Gaussian noise. In applications of robotics, shot noise may arise in the the form of massive proprioceptive measurement errors, large disturbances due to obstacle collisions, or natural external factors in the environment such as a large gust of wind. Despite this prevalence, there are little to no analytical controller or observer synthesis procedures dedicated towards robustness against shot noise perturbations. In fact, for any non-Gaussian noise processes, machine-learning-based methods~\cite{han16,shi20,tsukamoto2020neural} are often the go-to methods used to learn the model from scratch, but a downside of these approaches are the massive amounts of time and training data they consume. 

On the other hand, in the field of applied mathematics, there is an abundance of literature~\cite{applebaum09_book,oksendal_stoch_ctrl} which provides many theoretical results on Poisson random measures and jump-diffusion systems, which are relevant to addressing the controller and observer design procedures for shot-noise stochastic systems~\eqn{gen_shot_sde}. A particularly useful theoretical result known as the L\'{e}vy-Khintchine Decomposition Theorem, stated formally in Theorem 1.6 of~\cite{watson16}, Theorem 2.7 of~\cite{bass09}, or Theorem 1.2.14 of~\cite{applebaum09}, 
% which describes L\'{e}vy distributions as being weak limits of the convolution of Brownian motion processes and compound Poisson processes. This 
leads to a direct extension of~\eqn{gen_shot_sde} by uniting it with the Gaussian white noise case, yielding stochastic systems of the form:
\begin{align}\label{eq:gen_levy_sde}
    d\xvect(t) = f(t, \xvect)dt + \sigma(t,\xvect)dW(t) + \int_{\Rbb^{\ell}}\xi(t, \xvect, \zvect)N(dt,d\zvect)
\end{align}
This affine decomposition of \textit{bounded-measure L\'{e}vy noise processes} allows us to combine the controller and observer design procedures of~\eqn{gen_white_sde} and~\eqn{gen_shot_sde} to obtain a design procedure for~\eqn{gen_levy_sde}. 
% We note that L\'{e}vy noise processes are much more general than the additive form described here, but for sake of simpler terminology,

In this paper, we are motivated by two factors:
\begin{enumerate}
	\item the lack of discussion surrounding methodological control synthesis procedures for stochastic systems involving shot and L\'{e}vy noise processes
	\item the often inefficient way in which learning-based approaches are applied to any stochastic system perturbed by general non-Gaussian noise
\end{enumerate}
We take a step towards addressing both issues by proposing a \textbf{hierarchical controller synthesis} method composed of the following two parts:
\begin{enumerate}
	\item a primitive \textbf{pattern-learning component} which recognizes specific state sequences and stores in memory the corresponding control action applied to the system when a sequence has occurred
	\item a \textbf{modulation control component} which computes the control action to be taken when a specific state sequence is first encountered
\end{enumerate}
We note that in applications such as fault-tolerance, an independent lower-level controller can be used for the majority of the time, and the two-part controller is only applied when the system is destabilized. This requires some form of general stability analysis to determine a bound on when this two-part controller should be applied; in particular, we demonstrate how to determine this bound using the incremental stability analysis approach described in our previous paper~\cite{han21tac}. Moreover, most traditional methodological control design approaches do not include the first part, and we motivate our choice of its inclusion in our proposed method as follows. In applications where the magnitude of the jumps in the impulsive part of the noise process can be grouped into discrete values, there is potential for the system to observe re-occurring sequences. Hence, there is no need to devote time and computational energy to redundantly compute a control action for a scenario that has been observed before. We leverage results from the well-known theory of renewal processes~\cite{ross96book,ross06book,cox62book,cox65book} and the problem of ``pattern occurrence'' to obtain criteria, such as the expected time that elapses between two consecutive instances of the same pattern, to aid in designing the learning component. 
% The second part of the control policy computes the action to be taken when a pattern of  system 
% begins to grow more and more unstable. This requires explicit formulation of the problem as an optimal control problem with corresponding performance objectives. One notable method of Hamilton-Jacobi-Bellman-based control for jump-diffusion processes that we will use in this paper is known as the ``impulse control method'', and this approach has been used and described in several references such as~\cite{oksendal_stoch_ctrl,davis10}. Roughly speaking, the impulse control method which forces the state of the system down to some safer, lower-energy state once it exceeds a prespecified limit.
% % , so there is also relevance to applications in fault-tolerance control problems~\cite{yang20}.
We illustrate the proposed scheme using two particular case studies: 
\begin{enumerate}
    \item \textbf{fault-tolerance control}: we are inspired by fault-tolerance control applications such as spacecraft control~\cite{franchi18,wander13,kolcio16} to investigate a simple system which uses the two-part controller as a means of fault-tolerance to steer the system back to some predetermined, certified safety bound; when the trajectory is already within the safety bound in the absence of the random jump disturbances, an independent lower-level controller is used for stabilization.

    \item \textbf{congestion control}: the proposed two-part controller is used to optimally drive the state of the system towards increasingly lower energy states under a certain set of constraints. We consider the specific case study of controlling vehicle traffic at an intersection.
\end{enumerate}

\subsection{Paper Outline}\label{subsec:outline}
We begin in~\sec{learning_part} by laying out the renewal theory background relevant to the first part of the controller synthesis problem for shot noise stochstic systems~\eqn{gen_shot_sde} and L\'{e}vy noise stochastic systems~\eqn{gen_levy_sde}. 
% discusses three important extensions to renewal processes: 1) renewal reward processes, 2) delayed renewal processes, and 3) regenerative reward processes, all three of which become important for the formulation of the controller synthesis problem within the context of renewal theory.
The subsection~\ref{subsec:pattern_occurrence} then formally establishes the so-called pattern occurrence problem, which derives a way to compute the expected time between two consecutive occurrences of a specific pattern sequence in a string of renewals. Section~\ref{subsec:renewal_background} then provides an important unifying limiting theorem, called Blackwell's theorem, that is satisfied by all renewal processes and their corresponding extensions. 
% Finally, we take a detour in this section to connect the pattern occurrence problem to random walk theory, which can also be used to determine the expected time it takes the system to reach the boundary(ies) imposed on the system state, upon which the action determined by the second part of the hierarchical controller takes over.
% the impulse control method with random walk theory, which can be used to determine the expected time it takes the system to reach the boundary(ies) imposed on the system state, upon which the impulse control mechanism takes over.
Then~\sec{modulation_part} introduces the impulse control method to be employed in designing the actual action for the second part of the stochastic controller synthesis problem. We begin in Section~\ref{subsec:stoch_background} with a brief review of stochastic processes theory, the variations of It\^{o}'s formula, and the Poisson random measure in order to concretely establish the meaning and notation of the math we will be using for shot noise stochstic systems~\eqn{gen_shot_sde} and L\'{e}vy noise stochastic systems~\eqn{gen_levy_sde}, since these concepts may be unfamiliar to the reader. We then introduce the specific type of control known as impulse control in Section~\ref{subsec:impulse_control}, which will be used for the case studies of~\sec{case_studies}. Finally, in~\sec{case_studies}, we illustrate the control process on two sample case studies, the simple 1D linear stochastic process perturbed by L\'{e}vy noise, and the vehicle traffic at an intersection used to intuitively motivate our paper throughout the Introduction.

%%%%%%%%%%%%%%%%%%%%%%%%%%%%%%%%%%%%%%%%%%%%%%%%%%%%%%%%%%%%%%%%%%%%%%%
%% MOVED TO IMPULSE STUFF
% \section{Stochastic Processes Review}\label{sec:stoch_background}
% \import{arxiv_sections/}{ito_formula_review.tex}

%%%%%%%%%%%%%%%%%%%%%%%%%%%%%%%%%%%%%%%%%%%%%%%%%%%%%%%%%%%%%%%%%%%%%%%
\section{The Pattern-Learning Component}\label{sec:learning_part}

In this section, we discuss the first part of the controller synthesis procedure: the primitive learning component to recognize previously-occurred states so that the corresponding optimal control action can be recycled. To achieve this, we borrow elements from \textit{renewal theory}, and in particular, the \textit{pattern occurrence problem} to compute the expected time between consecutive patterns of renewals. Before we introduce the pattern occurrence problem in Section~\ref{subsec:pattern_occurrence}, we lay out some definitions and theoretical foundations at the core of renewal theory relevant for our discussion of stochastic control design in Section~\ref{subsec:renewal_background}. Most of these definitions have been presented in standard random processes references such as~\cite{ross96book} and~\cite{ross06book}, but we discuss them here to ensure that the material is self-contained. More importantly, the form of the theorems expressed in~\cite{ross96book} or~\cite{ross06book} are written for a general context, but we adapt the discussion for what is pertinent for the paper's proposed controller synthesis approach. For a more comprehensive treatment of our adapted discussion, we refer to 
% the extended manuscript~\cite{han21extended} and
our textbook preprint~\cite{han21book}.
% we limit discussion of the renewal theory background to what is relevant to the paper; additional mathematical results can be found in standard references for probability and random processes such as~\cite{ross06book,ross10book}. In particular, we establish and describe the ``'', which serves as the core to the design of the learning component, .
% and asks the following question: what is the expected time in between consecutive occurrences of a specific pattern of interest? 
% Finally, in Section~\ref{subsec:random_walks}, we address how this question can be solved using an alternative approach with random walks.

\subsection{Basic Definitions and Results from Renewal Theory}\label{subsec:renewal_background}
\begin{definition}[Counting Process]
    Suppose that we have a stochastic process $\{N(t), t\geq 0\}$ which satisfies the following conditions:
    \begin{itemize}
        \item the process takes nonnegative, integer values: $N(t) \geq 0$ and $N(t) \in \Zbb$
        \item the process is non-decreasing with respect to time: if $s\leq t$ then $N(s) \leq N(t)$
    \end{itemize}
    Then $N(t)$ is referred to as a \textit{counting process}.
\end{definition}
% Some examples of counting processes are standard Poisson processes

\begin{definition}[Renewal Process]
    Let $\{N(t), t\geq 0\}$ be a counting process and let $A_n := T_n - T_{n-1}$ be the interarrival times, where $T_n$ denotes the time of the $n$th arrival time. If $\{A_i\}_{i=1}^{\infty}$ is iid, then $N(t)$ is a called a \textit{renewal process}.
\end{definition}

Note that
\begin{align}\label{eq:dual_probs}
    \Pbb(N(t) = n) = \Pbb(N(t) \geq n) - \Pbb(N(t) \geq n+1) = \Pbb(T_n \leq t) - \Pbb(T_{n+1} \leq t)
\end{align}

There are numerous extensions to renewal processes which can be made in order to model a broader class of real-world phenomena. To model aspects of our controller synthesis procedure, we look at three specific extensions: 1) renewal reward processes, 2) delayed renewal processes, and 3) regenerative reward processes. Analysis of these three types of renewal processes requires using the following well-known result.

\begin{lemma}[Wald's Equation]\label{lem:wald}
    Let $\{A_i\}_{i=1}^{\infty}$ be iid sequence such that $\Ebb[A_i] := \Ebb[A]<\infty$ for all $i$, and let $N$ be a stopping time such that $\Ebb[N] < \infty$. Then the following equality holds:
    \begin{align}\label{eq:wald}
        \Ebb\left[ \sum\limits_{n=1}^N A_n\right] = \Ebb[N]\Ebb[A]
    \end{align}
\end{lemma}

The proof of this elementary result is straightforward and has been deferred to~\cite{han21book} to keep the content of the paper focused. Instead, we present and prove a more crucial result which examines an important limiting property for renewal processes, namely its time-average renewal rate. We will see later in Blackwell's theorem the three corresponding extensions of the time-average renewal rate for each of the three types of renewal processes mentioned above.
% \begin{proof}
%     Define the indicator random variable
%     \begin{align*}
%         I_n := \begin{cases}
%             1 &\text{ if } n \leq N\\
%             0 & \text{ else}
%         \end{cases}
%     \end{align*}
%     We can rewrite the left side of~\eqn{wald} as follows
%     \begin{align}\label{eq:wald_proof_1}
%         \Ebb\left[ \sum\limits_{n=1}^N A_n\right] = \Ebb\left[ \sum\limits_{n=1}^{\infty} A_nI_n \right] = \sum\limits_{n=1}^{\infty}\Ebb[A_nI_n] = \sum\limits_{n=1}^{\infty}\Ebb[A_n]\Ebb[I_n]
%     \end{align}
%     where the last equality follows from independence between $A_n$ and $I_n$. Continuing:
%     \begin{align*}
%         \eqn{wald_proof_1} = \Ebb[A]\sum\limits_{n=1}^{\infty}\Ebb[I_n] = \Ebb[A]\Ebb\left[ \sum\limits_{n=1}^{\infty} I_n\right] = \Ebb[A]\Ebb[N]
%     \end{align*}
%     which proves the desired relationship~\eqn{wald}.
% \end{proof}

\begin{theorem}\label{thm:elem_renewal}
    Let $N(t)$ be a renewal process with $\{A_i\}_{i=1}^{\infty}$ be iid sequence of interarrival times such that $\Ebb[A_i] := \Ebb[A] = \mu <\infty$ for all $i$. Then the \textit{time-average renewal rate} has the following limiting relationship:
        \begin{align}\label{eq:elem_renewal}
            \frac{m(t)}{t} \to \frac{1}{\mu} \text{ as } t \to \infty
        \end{align}
        where $m(t) := \Ebb[N(t)]$ is referred to as the \textit{mean process}.
\end{theorem}
% Before we proceed onwards with the proof,~\thm{elem_renewal} has a very intuitive interpretation which is easy to see in the well-known case of the Poisson process, in which the interarrival times $A_n$ are distributed according to exponential random variables. Suppose vehicles arrive at an intersection at an average rate of $\lambda = 3$ per hour. This implies that the average time between two consecutive vehicle arrivals is $1/\lambda = 1/3$ hour, or $20$ min. Hence, with time period $t$ denoting one hour, we have that $m(t)/t = 3$, and $1/\mu = 1/(1/3) = 3$, which verifies~\eqn{elem_renewal}.
\begin{proof}[Proof of~\thm{elem_renewal}.]
    We carry out this proof through two parts, we show that $\lim_{t\to\infty} \frac{m(t)}{t} \geq \frac{1}{\mu}$, then show that $\lim_{t\to\infty} \frac{m(t)}{t} \leq \frac{1}{\mu}$.

        Note that $T_{N(t) + 1}$ denotes the time of the first renewal after time $t$, which can alternatively be represented as
        \begin{align}\label{eq:elem_renew_proof_1}
            T_{N(t) + 1} = t + \Delta T_{N(t)}
        \end{align}
        where we refer to $\Delta T_{N(t)}$ as the ``excess'' time from $t$ until the next renewal. 

        By Wald's equation, note that
        \begin{align*}
            \Ebb[T_{N(t) + 1}] = \Ebb\left[ \sum\limits_{n=1}^{N(t)+1} A_n\right] = \Ebb[A]\Ebb[N(t) + 1] = \mu(m(t) + 1)
        \end{align*}
        which, in combination with~\eqn{elem_renew_proof_1}, yields
        \begin{align}\label{eq:elem_renew_proof_2}
            \mu(m(t) + 1) = t + \Ebb[\Delta T_{N(t)}] \ \Longrightarrow \ \frac{m(t)}{t} + \frac{1}{t} = \frac{1}{\mu} + \frac{\Ebb[\Delta T_{N(t)}]}{t\mu}
        \end{align}
        where the second equality follows from dividing through by $t\mu$. Since $\Delta T_{N(t)} \geq 0$, it follows that
        \begin{align}\label{eq:elem_renew_proof_3}
            \frac{m(t)}{t} \geq \frac{1}{\mu} - \frac{1}{t} \ \Longrightarrow \ \lim_{t\to\infty} \frac{m(t)}{t} \geq \frac{1}{\mu}
        \end{align}

        To prove the other half, suppose there exists a value $C < \infty$ such that $\Pbb(A_i < C) = 1$. This implies that $\Delta T_{N(t)} < C$, and so~\eqn{elem_renew_proof_2} implies
        \begin{align}\label{eq:elem_renew_proof_4}
            \frac{m(t)}{t} \leq \frac{1}{\mu} + \frac{C}{t\mu} - \frac{1}{t} \ \Longrightarrow \ \lim_{t\to\infty} \frac{m(t)}{t} \leq \frac{1}{\mu}
        \end{align}

        Thus, when the interarrival times are bounded,~\thm{elem_renewal} holds. In the case where they are unbounded, again fix $C > 0$, and define $\{N_C(t), t \geq 0\}$ to be the renewal process with interarrival times $\min(A_n, C), n \geq 1$. Since $\min(A_n, C) \leq A_n$ for all $n \geq 0$, $N_C(t) \geq N(t)$ for all $t \geq 0$ since the interarrival times are shorter. Consequently:
        \begin{align}\label{eq:elem_renew_proof_5}
            \lim_{t\to\infty} \frac{\Ebb[N(t)]}{t} \leq \lim_{t\to\infty} \frac{\Ebb[N_C(t)]}{t} = \frac{1}{\Ebb[\min(A_n, C)]}
        \end{align}
        where the second equality follows from the first case, where the interarrival times are bounded. Since $\lim_{C\to\infty} \Ebb[\min(A_n, C)] = \Ebb[A_n] = \mu$,~\eqn{elem_renew_proof_5} becomes
        \begin{align*}
            \lim_{t\to\infty} \frac{m(t)}{t} \leq \frac{1}{\mu}
        \end{align*}

        The two cases together, in combination with~\eqn{elem_renew_proof_3} yields the desired result~\eqn{elem_renewal}.
\end{proof}

Three natural extensions of renewal processes are described in the definition below.
% we can add a few extensions so that our controller synthesis problem can be modeled more accurately via renewal processes.

\begin{definition}\label{def:renewal_extensions}
    Let $\{N(t), t\geq 0\}$ be a renewal process with state space $\Zbb^{+}$ and interarrival times $\{A_i\}_{i=1}^{\infty}$.
    \begin{enumerate}
        \item Let $\{R_n\}_{n=1}^{\infty}$ be an iid sequences of random variables such that the $n$th renewal of $N(t)$ obtains a reward of $R_n$. Then $N(t)$ is said to be a \textit{renewal reward process}.

        \item $N(t)$ is said to be a \textit{delayed renewal process} if the first interarrival time of the process is distributed differently from all future interarrival times.

        \item $N(t)$ is said to be a \textit{regenerative renewal process} if there exist times $\{T_i^{(0)}\}_{i=1}^{\infty}$ at which the process restarts itself with probability $1$, i.e., $\Pbb(N(T_i^{(0)})) = 1$.
    \end{enumerate}
\end{definition}

Renewal processes and their extensions from~\defin{renewal_extensions} can be used to model elements of the case studies we investigate in~\sec{case_studies}. For instance, within the context of the congestion control problem, the cost that the system incurs by being at a certain level of congestion can be assigned a value using an appropriately-chosen reward function. In the vehicle traffic control problem, the reward of a vehicle could denote the inverse of the time spent waiting at the intersection before being allowed to pass. Regenerative renewal processes can be used to model the fault-tolerance control problem since the system effectively ``restarts'' from within the certified safety bound after the modulation control component corrects for the destabilizing pattern. Moreover, practical reasons for modeling systems using delayed renewal processes may be due to the system needing some additional time to warm up at the start.
Now, we introduce an important unifying limiting theorem for the mean process of a renewal process over time, known as \textit{Blackwell's theorem}, for different types of renewal processes. We can view Blackwell's theorem as a further development to the basic result described in~\thm{elem_renewal}. Several versions of Blackwell's theorem have been described in~\cite{ross96book},~\cite{ross06book}, as well as in~\cite{ross10book}. Section 4.4 of~\cite{cox62book} and Chapter 9 of~\cite{cox65book} further generalizes Blackwell's theorem and discusses some properties of renewal processes by invoking Laplace transforms. However, we emphasize that the version of Blackwell's theorem in this paper is specialized to make it easier to apply to the proposed controller synthesis framework, and also to the case studies that are addressed in~\sec{case_studies}. For an extensive treatment of Blackwell's theorem and the properties of renewal processes with the additional random processes background needed to facilitate their discussion, we refer to our textbook preprint~\cite{han21book}.

\begin{definition}[Lattice]
    A nonnegative random variable $S$ with cdf $F$ is \textit{lattice} if there exists a $c \geq 0$ such that $S$ only takes on values which are integer multiples of $c$. 
    \begin{align*}
        \sum\limits_{n=0}^{\infty} \Pbb(S = nc) = 1
    \end{align*}
    The largest such $c$ in which this property holds is referred to as the \textit{period} of $S$.
\end{definition}

We overload the terminology of ``lattice'' to describe both the random variable $S$ and its distribution function $F$.

\begin{definition}[Ladder Variables]
    An \textit{ascending variable of ladder height} $S_n$ occurs at time $n$ if $S_n > \max\{0, S_1, \cdots, S_{n-1}\}$. Similarly, a \textit{descending variable of ladder height} $S_n$ occurs at time $n$ if $S_n < \max\{0, S_1, \cdots, S_{n-1}\}$.
\end{definition}

% Suppose there is an ascending ladder variable of height $S_n$ at time $n$. Then the next ladder variable occurs at time $n+m$ if $m$ is the first value such that
% \begin{align*}
%     S_{n+m} > S_n \ \Longleftrightarrow \ X_{n+1} + \cdots + X_{n+m} > 0
% \end{align*}
% But note that since $X_i$ are iid, the probability of this event happening is the same regardless of which timestep we start on.
% \begin{align}\label{eq:ladder_var_iid}
%     \Pbb(X_{n+1} + \cdots + X_{n+m} > 0) = \Pbb(X_1 + \cdots + X_m > 0)
% \end{align}

% Denote
% \begin{align*}
%     p_{+} := \Pbb(\exists n \text{ s.t. } S_n > 0), \ p_{-} := \Pbb(\exists n \text{ s.t. } S_n < 0)
% \end{align*}
% Then by~\eqn{ladder_var_iid}, there will be exactly $m$ ascending ladder variables with probability $p_{+}^m(1 - p_{+})$. Likewise, for descending ladder variables, the probability is $p_{-}^m(1 - p_{-})$.

% ELEMENTARY RENEWAL THEOREM IS MOVED NOW.

\begin{theorem}[Blackwell's Theorem]\label{thm:massive_blackwell}
    Let $\{A_i\}_{i=1}^{\infty}$ be iid sequence such that $\Ebb[A_i] := \Ebb[A] = \mu <\infty$ for all $i$, and let $F$ be the distribution function of $A_i$. Let $N(t)$ be the corresponding renewal process, with $m(t) := \Ebb[N(t)]$.
    \begin{enumerate}
        \item If $F$ is not lattice, then
        \begin{align}\label{eq:blackwell_nonlattice}
            m(t+a) - m(t) \to \frac{a}{\mu} \ \text{ as } \ t \to \infty
        \end{align}
        for all $a \geq 0$.

        \item If $F$ is lattice with period $c \geq 0$, then
        \begin{align}\label{eq:blackwell_lattice}
            \Ebb[N(nc)] \to \frac{c}{\mu} \ \text{ as } \ n \to \infty
        \end{align}
        Note that $N(nc)$ denotes the number of renewals which have occurred by time $nc$.
    \end{enumerate}
\end{theorem}

\begin{proof}[Proof of~\thm{massive_blackwell}]
    We will only prove the nonlattice case of Blackwell's theorem statement, since the lattice case follows from a similar argument. For easier notation, let us denote $g(a) := \lim_{t\to\infty}(m(t+a) - m(t))$. Then note that 
    \begin{align*}
        g(a+b) = \lim_{t\to\infty}(m(t+a+b) - m(t+a)) + \lim_{t\to\infty}(m(t+a) - m(t)) = g(b) + g(a)
    \end{align*}
    The form of $g$ which satisfies this equation is given by $g(a) = c\cdot a$, for some constant $c$. Now we want to show that $c = 1/\mu$. Consider the following successive increments of $a = 1$:
    \begin{align*}
        \Delta m_1 &:= m(1) - m(0)\\
        \Delta m_2 &:= m(2) - m(1)\\
        &\vdots\\
        \Delta m_n &:= m(n) - m(n-1)
    \end{align*}
    Note that $\lim_{n\to\infty} \Delta m(n) = c$.

    On one hand, by the law of large numbers:
    \begin{align}\label{eq:blackwell_proof_1}
        \lim_{n\to\infty}\frac{1}{n}\sum\limits_{k=1}^n \Delta m(k) = c
    \end{align}
    On the other hand, by~\thm{elem_renewal}:
    \begin{align}\label{eq:blackwell_proof_2}
        \frac{1}{n}\sum\limits_{k=1}^n \Delta m(k) = \frac{m(n)}{n} \to \frac{1}{\mu} \qquad \text{ as } n\to\infty
    \end{align}
    Combining~\eqn{blackwell_proof_1} together with~\eqn{blackwell_proof_2}, we have that $c = 1/\mu$. This concludes the proof of the first part of the theorem. 
\end{proof}

\begin{corollary}[Extensions to Blackwell's Theorem]\label{coro:blackwell_extensions}
    %%% 5/7/2021 LECTURE
    There are also two straightforward extensions to Blackwell's theorem. The first 
    % There is also a straightforward extensions to Blackwell's theorem 
    is in the context of renewal reward processes. Suppose the renewal process $\{N(t), t \geq 0\}$ has iid rewards $\{R_i\}_{i=1}^{\infty}$. Assuming the distribution of the cycle of a renewal reward process is not lattice, then
    \begin{align}\label{eq:blackwell_reward}
        \Ebb[R(t,t+a)] \to \frac{a\Ebb[R]}{\Ebb[T]}
    \end{align}
    where $T$ is the length of a cycle of the renewal reward process.

    %%% 5/20/2021 MATERIAL
    % \label{thm:blackwell_line}
    The second extension is with respect to \textit{random walks}, which are set up as follows. Let $X_1, X_2, \cdots$ be iid with mean $\mu := \Ebb[X] < \infty$, and let $S_n := \sum_{i=1}^{\infty}X_i$. Typically, the random walk $S_n$ is started at $X_0 = S_0 = 0$, and we assume this implicitly throughout unless stated otherwise. Denote
    \begin{align*}
        U(t) := \sum\limits_{n=1}^{\infty} I_n \ \text{ where } \ I_n = \begin{cases} 1 &\text{ if } S_n \leq t\\ 0 &\text{ else}
        \end{cases}
    \end{align*}
    If $\mu > 0$ and $X_i$ are nonlattice, then
    \begin{align}\label{eq:blackwell_rw_line}
        \frac{u(t+a) - u(t)}{t} \to \frac{a}{\mu} \ \text{ as } \ t\to\infty \ \forall a > 0
    \end{align}
\end{corollary}

% \violet{PUSH TO BOOK VERSION?\\
% \begin{example}[Poisson Case]\label{ex:poisson_elem}

% Blackwell's theorem also admits an intuitive interpretation. 
% % The exponential distribution is a lattice distribution with period $c = 1$. 
% Suppose vehicles arrive at an intersection at an average rate of $\lambda = 3$ per hour, and the time period of consideration is $5$ hours. With a rate of $\lambda = 3$, it is easy to expect that the average number of vehicles observed is $3\cdot 5 = 15$. This corresponds to the statement of the theorem with $a = 5$.

% Blackwell's Theorem for renewal reward processes also admits an intuitive interpretation. Suppose that vehicles enter a parking lot which charges different costs to park depending on the vehicle's characteristics. Let the values of the costs range across three discrete values $R_n \in\{10,20,30\}$. Suppose that $\lambda=3$ vehicles arrive within a time range of one hour, with cost random variables $R_1, R_2, R_3$. Then the average cost among the three vehicles is computed as $\Ebb[R_1 + R_2 + R_3]/3 = 20$. In addition, for a time interval of 5 hours, we multiply this amount by $5$: the average reward accumulated over a timespan of $5$ hours is then given by $100$.
% % \end{example}
% % }

\begin{proof}[Proof of~\cor{blackwell_extensions}]
    Next, for the extension of Blackwell's theorem pertaining to renewal reward processes, note that 
    \begin{align*}
        \Ebb[R(t,t+a)] &:= \Ebb\left[ \sum\limits_{n=0}^{N(t+a)} R_n - \sum\limits_{n=0}^{N(t)} R_n\right]\\
        &= \left(\Ebb[N(t+a)] - \Ebb[N(t)]\right)\Ebb[R] \ \text{ by Wald's equation}\\
        &= (m(t+a) - m(t))\Ebb[R] \ \text{ by } m(t) \text{ definition}\\
        &\to \frac{a}{\Ebb[T]}\cdot\Ebb[R] \ \text{ as } t \to\infty
    \end{align*}
    where the limit follows from~\eqn{blackwell_nonlattice}.

    Finally, to prove Blackwell's result pertaining to random walks, we define a renewal process using successive ascending ladder heights as the renewals, and denote $Y(t)$ to be the excess height of the renewal process past height $t > 0$. Then $S_n = Y(t) + t$ is the first value of the random walk that exceeds $t$. See~\fig{blackwell_line_excess} for visualization. Hence, the primary difference between Blackwell's Theorem for Random Walks and the original Blackwell's Theorem or~\thm{elem_renewal} is that the renewal process has renewals which are dependent upon the previous renewals.
    % , by nature of ascending ladder variables.

    \begin{figure}
        \begin{center}
            \includegraphics[width=0.6\columnwidth]{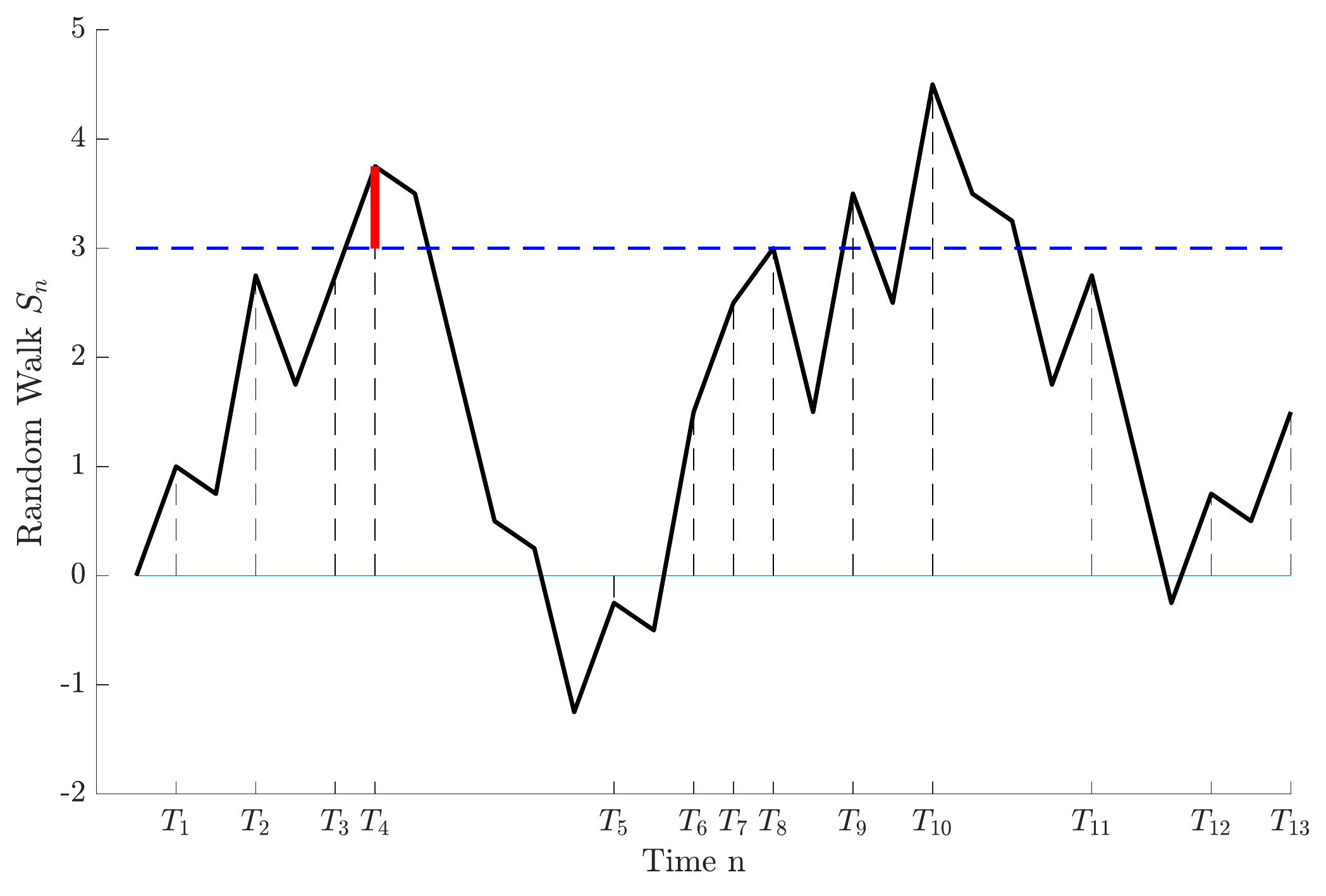}
        \end{center}
        \caption{The ascending ladder height renewal process with the excess at height $c = 3$. The excess $Y(t)$ is the length of the thick red line.}
        \label{fig:blackwell_line_excess}
    \end{figure}  

    The proof of this theorem adheres very closely to the technique of the proof for the original Blackwell's theorem~\eqn{blackwell_nonlattice}. Define
    \begin{align*}
        % g(Y(c)) &:= \Ebb\left[ U(c+a) - U(c) | Y(c)\right]
        h(a) &:= \lim_{t\to\infty} \left( u(t+a) - u(a)\right)
    \end{align*}
    Then we can show that $h(a+b) = h(a) + h(b)$ using the same logic as in the proof of~\eqn{blackwell_nonlattice}. The solution to such an equation is then given by $h(a) = \theta a$ for some constant $\theta$ to be determined.

    Let $\tau(t) := \min\{n | S_n > t\}$. We look at two cases of $X_i$, just as in the proof of~\eqn{blackwell_nonlattice}.

    First, if there exists $M > 0$ such that $X_i \leq M$ for all $i$, then
    \begin{align*}
        &t < \sum\limits_{i=1}^{\tau(t)} X_i < t + M \ \ 
        % \Longrightarrow \ &t < \Ebb[\tau(t)]\Ebb[X_i] < t + M \ \text{ by Wald}\\
        \Longrightarrow \ \ \frac{1}{\mu} < \frac{\Ebb[\tau(t)]}{t} < \frac{1}{\mu}\left(1 + \frac{M}{t}\right)
    \end{align*}
    which implies that 
    \begin{align}\label{eq:blackwell_bounded_X}
        \lim_{t\to\infty} \frac{\Ebb[\tau(t)]}{t} = \frac{1}{\mu}
    \end{align}

    On the other hand, if $X_i$ are unbounded, denote
    \begin{align*}
        N^{*}(t) := |\{ n > \tau(t) | S_n \in (-\infty, t]\}|
    \end{align*}
    to be the number of times after exceeding $c$ once $S_n$ lands in $(-\infty, t]$.

    Then note that
    \begin{align}\label{eq:blackwell_unbounded_X_1}
        U(t) = |\{ n | S_n \leq t\}| = (\tau(t) - 1) + N^{*}(t)
    \end{align}
    where the $-1$ comes from excluding the time $S_n$ exceeded $t$.

    Note that given the value of $Y(t) = y$, the distribution of $U(t+a) - U(t)$ is independent of $t$. If the first point of the random walk past $c$ occurs at $y + t$, then the number of points in $(t,t+a)$ has the same distribution as the number of points in $(0,a)$ given the first positive value of the random walk occurs at $y$. 
    % See~\fig{time_idpt_renewal} for visualization.

    Thus,
    \begin{align*}
        \Ebb[N^{*}(t)] \leq \Ebb\left[ | \{ n > \tau(0) | S_n < 0\}|\right]
    \end{align*}

    % \begin{figure}
    %     \begin{center}
    %         \includegraphics[width=0.4\columnwidth]{figures/time_idpt_renewal}
    %     \end{center}
    %     \caption{Placeholder: Conditional distribution of points is independent regardless of height shift.}
    %     \label{fig:time_idpt_renewal}
    % \end{figure}

    Since $\mu > 0$, $\Ebb[\tau(0)] < \infty$. At time $\tau(0)$, there is probability $1 - p_{-} > 0$ such that $S_n > S_{\tau(0)}$ for all $n > \tau(0)$. Otherwise, if there is such an $n$ where $S_n < S_{\tau(0)}$, then the expected additional time $m > 0$ such that $S_{n+m} > 0$ is finite since $\mu > 0$. From time $n + m$, there is again a probability of $1 - p_{-}$ $S_{n + k} > S_{\tau(0)}$ for all $k > m$. Thus:
    \begin{align*}
        &(1 - p_{-})\Ebb\left[ | \{ n > \tau(0) | S_n < 0\}|\right] \leq \Ebb[\tau(0) | X_1 < 0]\\
        \Longrightarrow \ \ &\Ebb\left[ | \{ n > \tau(0) | S_n < 0\}|\right] \leq \frac{\Ebb[\tau(0) | X_1 < 0]}{1 - p_{-}} < \infty
    \end{align*}
    This shows that $\Ebb[N^{*}(t)] < \infty$, and combined with~\eqn{blackwell_unbounded_X_1}, yields
    \begin{align}\label{eq:blackwell_unbounded_X}
        \lim_{t\to\infty} \frac{\Ebb[\tau(t)]}{t} = \lim_{t\to\infty} \frac{u(t)}{t}
    \end{align}

    Finally, to conclude the result from both~\eqn{blackwell_bounded_X} and~\eqn{blackwell_unbounded_X}, note that $u(1 + a) - u(a) \to \theta$ as $a \to \infty$. This implies
    \begin{align*}
        \frac{u(n+1) - u(1)}{n} = \frac{1}{n}\sum\limits_{a=1}^n u(1+a) - u(a) \to \theta \ \text{ as } \ n\to\infty
    \end{align*}
    Indeed, $\theta = 1/\mu$ and this concludes the proof.
\end{proof}

%%%%%%%%%%%%%%%%%%%%%%%%%%%%%%%%%%%%%%%%%%%%%%%%%%%%%%%
%%%%%%%%%%%%%%%%%%%%%%%%%%%%%%%%%%%%%%%%%%%%%%%%%%%%%%%

Now that we have discussed~\thm{elem_renewal} and Blackwell's theorem, we are ready to consider some other important properties of reward, delayed, and regenerative renewal processes, i.e. the arrival time of the last renewal before a certain time, or the mean number of renewals that have arrived by a certain time. This requires a result related to taking the integral with respect to the mean process of the renewal process. Before we present the result, we define Riemann-integrable functions within the context of renewal theory.

%% KEY RENEWAL THEOREM
% \violet{
\begin{definition}[Riemann-Integrable]
    Let $f: \Rbb^{+}\to\Rbb$ such that
    \begin{align*}
        \underline{f}_n(a) \leq f(t) \leq \overline{f}_n(a) \ \text{ for } \ t\in[(n-1)a, na]
    \end{align*}
    where $\underline{f}_n$ is the lower Riemann sum and $\overline{f}_n$ is the upper Riemann sum. We say that $f$ is \textit{Riemann-integrable} if
    \begin{itemize}
        \item $\sum\limits_{n=1}^{\infty} \underline{f}_n(a)$ and $\sum\limits_{n=1}^{\infty} \overline{f}_n(a)$ are finite for all $a > 0$
        \item $\lim_{a\to 0} a\sum\limits_{n=1}^{\infty} \underline{f}_n(a) = \lim_{a\to 0} a\sum\limits_{n=1}^{\infty} \overline{f}_n(a)$
    \end{itemize}
    Furthermore, a sufficient condition for $f$ to be Riemann integrable is that 1) $f(t) \geq 0$ for all $t\geq 0$, 2) $f(t)$ is nonincreasing, 3) $\int_0^{\infty} f(t)dt < \infty$.
\end{definition}
% }

\begin{theorem}\label{thm:key_renewal}
    Let $\{ N(t),t\geq 0\}$ be a renewal process with interarrival times $\{A_i\}_{i=1}^{\infty}$ being an iid sequence with distribution function $F$ (i.e., $F(t) := \Pbb(A_i = t)$) such that $\Ebb[A_i] := \Ebb[A] = \mu <\infty$ for all $i$. In addition, let $f: \Rbb^{+}\to\Rbb$ be a Riemann-integrable function. Then the following equality holds:
    \begin{align*}
        \lim_{t\to\infty} \int_0^t f(t-s) dm(s) = \frac{1}{\mu}\int_0^{\infty} f(s)ds
    \end{align*}
    where
    \begin{align*}
        m(x) := \Ebb[N(t)] = \sum\limits_{n=1}^{\infty} F_n(x), \qquad \mu := \int_0^{\infty} \overline{F}(t)dt
    \end{align*}
    where $F_n$ is the distribution of arrival time $T_n$ (which is the $n$-fold convolution of interarrival distribution $F$), and $\overline{F}(t) := \Pbb(A_i > t)$ for all $i$.    
\end{theorem}

The importance of~\thm{key_renewal} arises in computing the limiting value of some probability or expectation-like function $g(t)$ related to the renewal process. This yields an equation of the form
\begin{align*}
    g(t) = h(t) + \int_0^t h(t-s) dm(s)
\end{align*}
for some Riemann-integrable function $h$. One specific function is the distribution of $T_{N(t)}$, the arrival time of the last renewal before time $t$.
\begin{align*}
    \Pbb(T_{N(t)} \leq s) &= \sum\limits_{n=0}^{\infty} \Pbb(T_n \leq s, N(t) \leq n)\\
    % &= \sum\limits_{n=0}^{\infty} \Pbb(T_n \leq s, T_{n+1} > t) \ \text{ since } \{N(t) \leq n\} \Longleftrightarrow \{T_{n+1}>t\}\\
    &= \Pbb(T_0 \leq s, T_1 > t) + \sum\limits_{n=1}^{\infty} \Pbb(T_n \leq s, T_{n+1} > t) \ \text{ since } \{N(t) \leq n\} \Longleftrightarrow \{T_{n+1}>t\}\\
    &= \Pbb(T_1 > t) + \sum\limits_{n=1}^{\infty} \Pbb(T_n \leq s, T_{n+1} > t) \ \text{ since } T_0 := 0 \leq s \text{ always}\\
    % &= \overline{F}(t) + \sum\limits_{n=1}^{\infty} \Pbb(T_n \leq s, T_{n+1} > t)\\
    &= \overline{F}(t) + \sum\limits_{n=1}^{\infty} \int_0^{\infty} \Pbb(T_n \leq s, T_{n+1} > t | T_n \leq r) dF_n(r)\\
    &= \overline{F}(t) + \sum\limits_{n=1}^{\infty} \int_0^s \Pbb( T_{n+1} - T_n > t - r) dF_n(r)\\
    &= \overline{F}(t) + \int_0^s \overline{F}(t-r) d\left(\sum\limits_{n=1}^{\infty} F_n(r)\right)\\
    &= \overline{F}(t) +  \int_0^s \overline{F}(t-r) dm(r)
\end{align*}
Hence:
\begin{align}\label{eq:key_renewal_TNt}
    \Pbb(T_{N(t)} \leq s) &= \overline{F}(t) +  \int_0^s \overline{F}(t-r) dm(r)
\end{align}

Now we are ready to describe the properties of reward, delayed, and regenerative renewal processes, respectively.

% \subsection{Renewal Reward Processes}\label{subsec:renewal_reward}

\begin{lemma}[Property of a Renewal Reward Process]\label{lem:renewal_reward}
    Let $N(t)$ be a renewal reward process with interarrival times $\{A_n\}_{n=1}^{\infty}$ and rewards $\{R_n\}_{n=1}^{\infty}$. Moreover, let $R(t)$ represent the total reward earned by time $t$:
    \begin{align*}
        R(t) := \sum\limits_{n=1}^{N(t)} R_n
    \end{align*}
    and denote $\nu := \Ebb[R_n], \mu := \Ebb[A_n]$ for all $n \geq 1$. Further suppose $\nu, \mu < \infty$. Then the following hold:
    \begin{subequations}
        \begin{align}
            \lim_{t\to\infty} \frac{R(t)}{t} &= \frac{\nu}{\mu} \ \text{ w.p. } 1\label{eq:renewal_1}\\
            \lim_{t\to\infty} \frac{\Ebb[R(t)]}{t} &= \frac{\nu}{\mu}\label{eq:renewal_2}
        \end{align}
    \end{subequations}
\end{lemma}

\begin{proof}[Proof of~\lem{renewal_reward}]
    Since the proof of~\eqn{renewal_2} follows similarly to the proof of~\eqn{renewal_1}, we will only prove~\eqn{renewal_1}. Write:
    \begin{align*}
        \frac{R(t)}{t} := \frac{1}{t}\sum_{n=1}^{N(t)}R_n = \left(\frac{1}{N(t)}\sum_{n=1}^{N(t)}R_n\right)\left(\frac{N(t)}{t}\right)
    \end{align*}
    By the strong law of large numbers, $(\sum_{n=1}^{N(t)}R_n)/N(t) \to \nu$ as $t\to\infty$, while $N(t)/t \to 1/\mu$ as $t\to\infty$ follows from~\eqn{elem_renewal}. The combination yields~\eqn{renewal_1}.
\end{proof}

% \subsection{Delayed Renewal Processes}\label{subsec:delayed_renewal}

\begin{lemma}[Properties of a Delayed Renewal Process] %PROBLEM 3.25
    % \blue{(PROBLEM 3.25, SEC. 3.4.2)}\\
    Consider a delayed renewal process $\{N_D(t), t\geq 0\}$, with first interarrival time distribution $A_1 \sim G$ with finite mean, i.e. $G(t) := \Pbb(T_1 < t)$, and successive interarrival time distribution $A_2,A_3,\cdots \sim F$, where $F$ is nonlattice with $\int s^2dF(s) < \infty$. Then the following properties hold:
    \begin{itemize}
        \item Denoting $m_D(t) := \Ebb[N_D(t)]$ to be the \textit{mean number of renewals} by time $t$:
        \begin{align}\label{eq:mean_delay_renewal}
            m_D(t) &= G(t) + \int_0^t \sum\limits_{n=1}^{\infty}F_n(t-s) dG(s)
        \end{align}

        \item Denoting $H_D(t) := t - T_{N_D(t)}$ to be the \textit{age} of the process by time $t$, we have $t\overline{G}(t) \to 0$ as $t\to\infty$ and
        \begin{align}\label{eq:age_delay_renewal}
            \Ebb[H_D(t)] \to \left(\int_0^{\infty}s^2 dF(s)\right)\left(2\int_0^{\infty}sdF(s)\right)^{-1} \ \text{ as } t\to\infty
        \end{align}
    \end{itemize}
\end{lemma}

\begin{proof}
    First, by definition of expectation and utilizing the equivalence in~\eqn{dual_probs}:
    \begin{align*}
        m_D(t) := \Ebb[N_D(t)] &= \sum\limits_{n=0}^{\infty} \Pbb(N_D(t) > n)\\
        &= \Pbb(N_D(t) > 0) + \sum\limits_{n=1}^{\infty} \Pbb(N_D(t) > n)\\
        % &= \Pbb(T_1 < t) + \sum\limits_{n=1}^{\infty} \Pbb(N_D(t) > n)\\
        &= G(t) + \sum\limits_{n=1}^{\infty} \int_0^t \Pbb(N_D(t) > n | T_1 \leq s) dG(s)\\
        % &= G(t) + \sum\limits_{n=1}^{\infty} \int_0^t \Pbb(T_{n+1} \leq t | T_1 \leq s) dG(s)\\
        % &= G(t) + \sum\limits_{n=1}^{\infty} \int_0^t \Pbb(T_n \leq t - s) dG(s)\\
        &= G(t) + \int_0^t \sum\limits_{n=1}^{\infty} F_n(t-s) dG(s)
    \end{align*}
    This proves~\eqn{mean_delay_renewal}.

    Second, we first have by the dominated convergence theorem
    \begin{align}\label{eq:dct_property_1}
        \lim_{t\to\infty} t\overline{G}(t) = \lim_{t\to\infty} t\int_t^{\infty} G(s)ds \leq \lim_{t\to\infty} \int_t^{\infty} sG(s)ds
    \end{align}
    Note that because the mean of $G$ is assumed to be finite, the integrand of~\eqn{dct_property_1} is finite. Hence, when taking $t\to\infty$, the value of the overall integral tends to 0 since the upper and lower limits converge to the same value.

    Now we can consider the age of the delayed renewal process. By conditioning on the value of $T_{N_D(t)}$ and substituting in~\eqn{key_renewal_TNt} (and taking care to ensure that the distribution of the first interarrival time is given by $G$, not $F$), we get
    \begin{align}\label{eq:mean_delay_renewal_1}
        \Ebb[H_D(t)] &= \Ebb[H_D(t) | T_{N_D(t)} = 0]\Pbb(T_{N_D(t)} = 0) + \int_0^t \Ebb[H_D(t) | T_{N_D(t)} = s]\Pbb(T_{N_D(t)} = s) ds\notag\\
        &= \Ebb[H_D(t) | T_{N_D(t)} = 0]\overline{G}(t) + \int_0^t \Ebb[H_D(t) | T_{N_D(t)} = s]\overline{F}(t-s)dm_D(s)\notag\\
        &= \Ebb[H_D(t) | A_1 > t]\overline{G}(t) + \int_0^t \Ebb[H_D(t) | T_{N_D(t)} = s]\overline{F}(t-s)dm_D(s), \ \text{ for } n \geq 2\notag\\
        &= t\overline{G}(t) + \int_0^t (t-s)\overline{F}(t-s)dm_D(s)
    \end{align}
    since note that the age is simply the time elapsed since the last renewal. Now, note that under the assumption that $G$ is finite, we showed above that $t\overline{G}(t)\to 0$ as $t\to\infty$, which implies that the first term of~\eqn{mean_delay_renewal_1} tends to $0$. We will thus ignore the first term in our analysis and focus primarily on the second term. Using~\thm{key_renewal}, 
    % and the fact that the equilibrium delayed renewal process has mean which converges to
    % \begin{align*}
    %     m_D(t) \to \frac{t}{\mu}
    % \end{align*}
    we get:
    \begin{align}\label{eq:mean_delay_renewal_2}
        \int_0^t (t-s)\overline{F}(t-s)dm_D(s) \to \frac{1}{\mu}\int_0^{\infty} s\overline{F}(s) ds
    \end{align}

    Note the relationship
    \begin{align*}
        \overline{F}(s) = \Pbb(Y > s) = \int_0^{\infty} \Pbb(Y=u) du
    \end{align*}

    Substituting, we get:
    \begin{align*}
        \eqn{mean_delay_renewal_2} &= \frac{1}{\mu}\int_0^{\infty} s \left(\int_s^{\infty} dF(u)\right)ds\\
        &= \frac{1}{\mu}\int_0^{\infty}\left(\int_0^u s ds\right)dF(u) \ \text{ via change of variables}\\
        &= \frac{1}{2\mu}\int_0^{\infty} u^2 dF(u) = \left(\int_0^{\infty} s^2 dF(s)\right)\left(2\int_0^{\infty} s dF(s)\right)^{-1}
    \end{align*}
    This proves~\eqn{age_delay_renewal}.
\end{proof}

% \subsection{Regenerative Renewal Processes}\label{subsec:regen_renewal}

\begin{lemma}[Property of a Regenerative Process]\label{lem:regenerative}
    % \blue{THEOREM 3.7.1}\\
    Denote $A^{(0)}$ to be the time length of a cycle for regenerative process $\{N(t),t\geq 0\}$, i.e. the duration of time between two consecutive time points at which the state of the renewal process is $0$. Denote $A_i^{(0)}$ to be the total amount of time spent in state $i$ in between two consecutive times at which the renewal process is in state $0$. Further suppose the distribution $F$ of a cycle has a density over some interval of time and $\Ebb[A_i^{(0)}] < \infty$ for all $i$. Then the following property holds:
    \begin{align}
        P_i := \lim_{t\to\infty} P(t) := \lim_{t\to\infty} \Pbb(N(t) = i) = \frac{\Ebb[A_i^{(0)}]}{\Ebb[A^{(0)}]}
    \end{align}
\end{lemma}

\begin{proof}
    Conditioning on the time of the last cycle before time $t$, and using~\eqn{key_renewal_TNt} yields:
    \begin{align}\label{eq:regenerative_proof_1}
        P(t) = \Pbb(N(t) = i | T_{N(t)} = 0)\overline{F}(t) + \int_0^t \Pbb(N(t) = i | T_{N(t)} = s)\overline{F}(t-s)dm(s)
    \end{align}
    where $m(t) := \Ebb[N(t)]$. We've seen before that
    \begin{align*}
        \Pbb(N(t) = i | T_{N(t)} = 0) &= \Pbb(N(t) = i | A_1^{(0)} > t)\\
        \Pbb(N(t) = i | T_{N(t)} = s) &= \Pbb(N(t) = i | A_1^{(0)} > t-s)
    \end{align*}
    and hence:
    \begin{align}
        \eqn{regenerative_proof_1} &= \Pbb(N(t) = i | A_1^{(0)} > t)\overline{F}(t) + \int_0^t \Pbb(N(t) = i | A_1^{(0)} > t-s)\overline{F}(t-s)dm(s)\notag\\
        &\to \frac{1}{\Ebb[A_1^{(0)}]}\int_0^{\infty} \Pbb(N(t) = i | A_1^{(0)} > s)\overline{F}(s)ds \ \text{ by~\thm{key_renewal}}
    \end{align}
    and this proves the result since the integral denotes exactly the amount of time within a cycle $A^{(0)}$ the system spends in state $i$.
\end{proof}

\subsection{The Pattern Occurrence Problem}\label{subsec:pattern_occurrence}
%%%%% VERSION OF PATTERN OCCURRENCE SECTION WITH NEW NON-ROSS EXAMPLE. FOR THE BOATLOAD OF COMMENTED OUT MATERIAL, SEE pattern_occurrence.tex.

With the renewal theory background established, we now consider the general problem of \textit{pattern occurrence}. We first discuss a couple motivational examples before describing a more general result. 

\begin{example}[Motivating Example: Expected Time]\label{ex:brute_force}
    Suppose that we are observing a sequence of iid binary random variables $X_1, X_2, \cdots$, each with distribution
    \begin{align*}
        X_i = \begin{cases}
            1 &\text{ with probability } \ p\\
            0 &\text{ with probability } \ q := 1-p
        \end{cases}
    \end{align*}
    Suppose we designate two strings $A_1$ and $A_2$ composed of $0$'s and $1$'s. One common problem of interest is the expected number of trials it takes to observe two consecutive instances of $A_1$ in the sequence $\{X_i\}_{i=1}^{\infty}$; likewise, for $A_2$. Denote $T_i$ to be the time in between two consecutive instances of $A_i$, where $i=1,2$.

    We choose for concreteness $A_1 := (1,0,1,0,\cdots,1,0)$ to be the length-$2n$ alternating sequence of $1$ and $0$, and $A_2 := (1,\cdots,1,0,\cdots,0)$ to be the length-$m$ string of all $1$'s followed by the length-$m$ string of all $0$'s. 

    We first compute $\Ebb[T_1]$. Note that if we were to treat each pair of $(1,0)$ as a single symbol, then we can view $A_1$ as a length-$n$ string composed entirely of $(1,0)$ pairs. Clearly, the probability of observing exactly $(1,0)$ is $pq$. Hence, the number of trials required to observe $(1,0)$ for the first time after the last occurrence of $A_1$ is distributed according to a Geometric random variable with parameter $pq$. We can now iteratively derive the expression for the expected time it takes to obtain a new string of $n$ pairs of $(1,0)$ after the last occurrence.
    % Note that if $(1,0)$ appears on the spot, we are done, because we can combine this new $(1,0)$ with the last $n-1$ occurrences of $(1,0)$ from the last instance of $A_1$ to create the new instance of $A_1$. However, if $(1,0)$ doesn't appear exactly, we need to restart as if we were observing $A_1$ from scratch. The expected time it takes to observe the first instance of $(1,0)$ is $1/(pq)$ due to the Geometric distribution. Once the first $(1,0)$ has been observed, the time it takes to observe the second instance of $(1,0)$ is also Geometrically distributed, but with parameter $p^2q^2$ because we must observe the two $(1,0)$'s consecutively; the probability of observing exactly the sequence $(1,0,1,0)$ is thus $p^2q^2$. Hence, we have that the mean number of trials to observe two consecutive $(1,0)$'s for the first time is $(1/(pq)) + (1/(pq)^2)$. 
    % Iterating this argument until $n$ consecutive $(1,0)$'s, we have that the expected number of trials in between consecutive instances of $A_1$ is:
    \begin{align}\label{eq:motive_success_time}
        \Ebb[T_1] = \sum\limits_{j=1}^n \frac{1}{(pq)^j}
    \end{align}

    The expression~\eqn{motive_success_time} can also be directly derived using a system of linear equations. Denote $E_{n,j}$ to be the expected number of trials needed to observe $n$ consecutive $(1,0)$'s given $j$ consecutive $(1,0)$'s, where $0\leq j\leq n$, have already been observed. Further denote $E_{n,j}^{(1)}$ to be the expected number of trials needed to observe $n$ consecutive $(1,0)$'s given $j$ consecutive $(1,0)$'s and an additional trial with outcome $1$ have been observed. We now have $n$ equations with these variables set up as follows:
    \begin{align*}
        E_{n,n-k} &= 1 + pE_{n,n-k}^{(1)} + qE_{n,0}\\
        E_{n,n-k}^{(1)} &= 1 + pE_{n,0}^{(1)} + qE_{n,n-k+1}, \ \ \forall 2\leq k\leq n\\
        E_{n,n} &= 1
    \end{align*}
    Solving this system and setting $\Ebb[T_1] = E_{n,0}$ yields the same expression as in~\eqn{motive_success_time}. 
    % For details on how these equations were derived, we refer the reader to~\cite{han21book}.

    % Note that we can simplify the two equations and remove the intermediate variables $E_{n,n-k}^{(1)}$, for $1\leq k\leq n-1$ via substitution:
    % \begin{align*}
    %     E_{n,n-k} = 1 + p + p^2E_{n,0}^{(1)} + pqE_{n,n-k+1} + qE_{n,0}
    % \end{align*}

    % We can then solve for the remaining variables by iterating from $k=n$ to $k=2$:
    % \begin{align*}
    %     E_{n,0}^{(1)} &= E_{n,1} + \frac{1}{q}\\
    %     E_{n,0} &= E_{n,1} + \frac{1}{q} + \frac{1}{p} = E_{n,1} + \frac{1}{pq}\\
    %     E_{n,1} &= E_{n,2} + \frac{1}{(pq)^2}\\
    %     &\vdots\\
    %     E_{n,k} &= E_{n,k+1} + \frac{1}{(pq)^k}
    % \end{align*}
    % and setting $\Ebb[T_1] = E_{n,0}$ yields the same expression as in~\eqn{motive_success_time}.

    We compute $\Ebb[T_2]$ using a similar approach. 
    % Unlike the pattern $A_1$, there is no subsequence from the end of the last instance of $A_2$ which we can reuse as a part of the next instance of $A_2$. We effectively need to count the number of trials as if we were starting the sequence from scratch. 
    We must first observe the $m$ consecutive $1$'s before considering the $m$ consecutive $0$'s. Note that the probability of observing exactly $m$ consecutive $1$'s is $p^m$; hence, the time it takes to observe $m$ consecutive $1$'s is Geometrically distributed with parameter $p^m$. Once $m$ $1$'s have been observed, the time it takes to observe $m$ consecutive $0$'s is also Geometrically distributed. Put together, the time to observe $A_2$ is Geometrically distributed with parameter $p^mq^m$. 
    % However, with the same argument used for $A_1$, the $m$ consecutive $0$'s must follow immediately after the $m$ consecutive $1$'s. 
    \begin{align}\label{eq:motive_failure_time}
        \Ebb[T_2] = \frac{1}{(pq)^m}
    \end{align}

    The expression~\eqn{motive_failure_time} can also be derived by solving a system of equations. Denote $E_{k,i}$ for $0\leq k\leq m$ and $i=0,1$ to be the expected number of trials needed to observe the full $A_2$ given we have observed the subsequence of $A_2$ up to the $k$th outcome of $i$. Then:
     % the equations are
    \begin{align*}
        E_{k,1} &= 1 + pE_{k+1,1} + qE_{0,1}, \ \ \forall \ 0\leq k\leq m-1\\
        E_{m,1} &= 1 + pE_{m,1} + qE_{1,0},\\
        E_{k,0} &= 1 + pE_{1,1} + qE_{k+1,0}, \ \ \forall \ 1\leq k\leq m\\
        E_{m,0} &= 1
    \end{align*}
    Solving this system and setting $\Ebb[T_2] = E_{0,1}$ yields the same expression as in~\eqn{motive_success_time}. For further details on the derivation for both $\Ebb[T_1]$ and $\Ebb[T_2]$, see~\cite{han21book}.

    % From the first $m$ equations, we get that:
    % \begin{align*}
    %     E_{1,1} = E_{m,1} + \frac{1}{p^2} + \cdots + \frac{1}{p^m} = E_{m,1} + \underbrace{\frac{p(1 - p^{m-2})}{qp^m}}_{=: K}
    % \end{align*}
    % and from the next $m+1$ equations, we get:
    % \begin{align*}
    %     E_{1,0} &= E_{m,1} - \frac{1}{q} = E_{1,1} - K - \frac{1}{q}\\
    %     E_{k,0} &= E_{1,1} - \frac{1}{q^{k-1}}\left( K + \frac{1}{q}\right) - \frac{1}{q^{k-1}} - \cdots - \frac{1}{q}, \ \ \forall \ 2\leq k\leq m
    % \end{align*}

    % In particular, substituting $k = m$ and using the boundary condition:
    % \begin{align*}
    %     1 = E_{m,0} &= E_{1,1} - \frac{1}{q^{m-1}}\left( K + \frac{1}{q}\right) - \frac{1}{q^{m-1}} - \cdots - \frac{1}{q}\\
    %     \Longrightarrow E_{0,1} &= \frac{1}{p} + 1 + \frac{1}{q^{m-1}}\left( \frac{p(1 - p^{m-2})}{qp^m} + \frac{1}{q}\right) + \frac{q(1 - q^{m-1})}{pq^m}
    % \end{align*}
    % and simplification of the expression yields the expression~\eqn{motive_failure_time}.
\end{example}

\begin{example}[Motivating Example: Probability]\label{ex:brute_force_ctd}
    Given a sequence of binary random variables $X_1, X_2, \cdots$, another problem of interest is the probability of observing one subsequence before another. 
    % We consider solving a generalization of a simple problem from~\cite{ross06book}. 
    Let $A_1$ be a string of $n$ $1$'s, $A_2$ be a string of $m$ $0$'s, and $A_3$ be a string of $r$ $2$'s, where $n, m, r\in\Nbb^{+}$. Suppose the values of $\{X_i\}_{i=1}^{\infty}$ are generated in the following way. Until one of the two sequences $A_1$ or $A_2$ have been observed:
    \begin{align*}
        X_i = \begin{cases}
            1 &\text{ with probability } \ p_1\\
            0 &\text{ with probability } \ q_1 := 1-p_1
        \end{cases}
    \end{align*}
    If $A_1$ is observed first, the $X_i$ take the following distribution until either $A_2$ or $A_3$ is observed:
    \begin{align*}
        X_i = \begin{cases}
            2 &\text{ with probability } \ p_2\\
            0 &\text{ with probability } \ q_2 := 1-p_2
        \end{cases}
    \end{align*}
    Otherwise, if $A_2$ is observed first, the $X_i$ take the following distribution until either $A_1$ or $A_3$ is observed:
    \begin{align*}
        X_i = \begin{cases}
            1 &\text{ with probability } \ p_3\\
            2 &\text{ with probability } \ q_3 := 1-p_3
        \end{cases}
    \end{align*}

    Given this setup, we are interested in the probability $P_{1,3}$ that $A_1$ is observed before $A_2$, and then $A_3$ is observed before $A_2$. Denote $\Ecal_1$ to be the event that $A_1$ is observed before $A_2$, and let $P_1 := \Pbb(\Ecal_1)$. Further denote $\Ecal_3$ to be the event that $A_3$ is observed before $A_2$ given $A_1$ has been observed first, and let $P_3 := \Pbb(\Ecal_3)$. Then by conditional probability, we can decompose the probability into two parts:
    \begin{align*}
        P_{1,3} = P_1P_3
    \end{align*}

    Now $P_1$ and $P_3$ can then be computed independently. We first compute $P_1$ by conditioning on the outcome of the first trial $X_1$. To that end, we denote $\Scal_1$ to be the event that $X_2=\cdots=X_n=1$ and $\Scal_0$ to be the event that $X_2=\cdots=X_m = 0$. Thus:
    \begin{align}\label{eq:motive_prob_eq1}
        P_1 := \Pbb(\Ecal) 
        % &= p_1\Pbb(\Ecal|X_1=1) + q_1\Pbb(\Ecal|X_1=0)\notag\\
        &= p_1\left(p_1^{n-1}\Pbb(\Ecal|X_1=1,\Scal_1) + (1 - p_1^{n-1})\Pbb(\Ecal|X_1=1,\Scal_1^{c})\right)\notag\\
        &\hskip1cm + q_1\left( q_1^{m-1}\Pbb(\Ecal|X_1=0,\Scal_0) + (1 - q_1^{m-1})\Pbb(\Ecal|X_1=0,\Scal_0^{c})\right)
    \end{align}
    Note that $\Pbb(\Ecal|X_1=1,\Scal_1) = 1$ and $\Pbb(\Ecal|X_1=0,\Scal_0) = 0$. Furthermore, $\Pbb(\Ecal|X_1=1,\Scal_1^{c}) = \Pbb(\Ecal|X_1=0)$ because $\Scal_1^{c}$ means that the outcome is $0$ for one of the trials among $X_2, \cdots, X_n$, after which we need to restart the count; the probability then becomes equivalent to $\Pbb(\Ecal|X_1=0)$. Likewise, $\Pbb(\Ecal|X_1=0,\Scal_0^{c}) = \Pbb(\Ecal|X_1=1)$. We then arrive at a system of two equations and two unknowns $\Pbb(\Ecal|X_1=0)$ and $\Pbb(\Ecal|X_1=1)$, which we can solve and substitute back into~\eqn{motive_prob_eq1}. We obtain an equivalent expression for $P_3$ via similar calculations. 
    % Note then we now have a system of two equations:
    % \begin{align*}
    %     \Pbb(\Ecal|X_1=1) = p_1^{n - 1} + (1 - p_1^{n - 1})\Pbb(\Ecal|X_1=0), \qquad \Pbb(\Ecal|X_1=0) = (1 - q_1^{m-1})\Pbb(\Ecal|X_1=1)
    % \end{align*}
    % which we can solve and substitute back into~\eqn{motive_prob_eq1}:
    % \begin{align}\label{eq:motive_prob}
    %     P_1 = \eqn{motive_prob_eq1} &= p_1\left(p_1^{n-1} + (1 - p_1^{n-1})\Pbb(\Ecal|X_1=0)\right) + q_1(1 - q_1^{m-1})\Pbb(\Ecal|X_1=1)\notag\\
    %     &= \frac{q_1^{m - 1}(1 - p_1^n)}{p_1^{n - 1} + q_1^{m - 1} - p_1^{n - 1}q_1^{m - 1}}
    % \end{align}
    % Note that $P_3$ can be computed using the same formula~\eqn{motive_prob}:
    % \begin{align*}
    %     P_3 = \frac{q_2^{m - 1}(1 - p_2^n)}{p_2^{n - 1} + q_2^{m - 1} - p_2^{n - 1}q_2^{m - 1}}
    % \end{align*}
    Hence:
    \begin{align}\label{eq:motive_prob}
        P_{1,3} = \left(\frac{p_1^{n - 1}(1 - q_1^m)}{q_1^{m - 1} + p_1^{n - 1} - q_1^{m - 1}p_1^{n - 1}}\right)\left(\frac{p_2^{r - 1}(1 - q_2^m)}{q_2^{m - 1} + p_2^{r - 1} - q_2^{m - 1}p_2^{r - 1}}\right)
    \end{align}

    % more application to the fault-tolerance switching problem. Conditioning state of system (X_i) on whether A_1 vs A2 shows up first.

    Now return back to the case from~\exa{brute_force} where $A_1 = (1,0,1,0,\cdots, 1,0)$ with length $2n$ and $A_2 = (1,\cdots,1,0,\cdots,0)$ with length $2m$. For the sake of simplicity, we will take the symmetric case of $p=q=1/2$. Again define $\Ecal$ to be the event where $A_1$ is observed before $A_2$, and let $P_1 := \Pbb(\Ecal)$. We can again compute $P_1$ by conditioning on the outcomes of trials, but in this case, both $A_1$ and $A_2$ start with $1$. Thus, we instead condition on the outcome of the first two trials $X_1$ and $X_2$. To that end, we denote $\Scal_1$ to be the event that $X_3=1,X_4=0,\cdots=X_{2n-1}=1,X_{2n}=0$ and $\Scal_2$ to be the event that $X_3=\cdots=X_m=1,X_{m+1}=\cdots=X_{2m}=0$. Thus:
    \begin{align*}
        P_1 := \Pbb(\Ecal) &= p\Pbb(\Ecal|X_1=1) + q\Pbb(\Ecal|X_1=0)\\
        &= p\left( p\Pbb(\Ecal|X_1=1,X_2=1) + q\Pbb(\Ecal|X_1=1,X_2=0)\right) + q\Pbb(\Ecal)
    \end{align*}
    where $\Pbb(\Ecal|X_1=0) = \Pbb(\Ecal)$ because neither $A_1$ nor $A_2$ start with $0$.

    Moreover, conditioning on $\Scal_1$ and $\Scal_2$, we get the following respective equations:
    % using the same logic as in the previous problem:
    \begin{align*}
        \Pbb(\Ecal|X_1=1,X_2=0) 
        % = p^{n-1}q^{n-1} + (1 - p^{n-1}q^{n-1})\Pbb(\Ecal|X_1=1,X_2=0,\Scal_1^c)\\
        % &= p^{n-1}q^{n-1} + (1 - p^{n-1}q^{n-1})\left[\frac{1}{2}\Pbb(\Ecal|X_1=1,X_2=0,\text{1 flipped to 0}) + \frac{1}{2}\Pbb(\Ecal|X_1=1,X_2=0,\text{0 flipped to 1})\right]\\
        &= p^{n-1}q^{n-1} + (1 - p^{n-1}q^{n-1})\left[\frac{1}{2}\Pbb(\Ecal) + \frac{1}{2}\Pbb(\Ecal|X_1=1,X_2=1)\right]\\
    % \end{align*}
    % and
    % \begin{align*}
        \Pbb(\Ecal|X_1=1,X_2=1)
        %  = (1 - p^{m-2}q^m)\Pbb(\Ecal|X_1=1,X_2=1,\Scal_2)\\
        % &= (1 - p^{m-2}q^m)\left[ \frac{m-2}{2m-2}\Pbb(\Ecal|X_1=1,X_2=1,\text{1 flipped to 0}) + \frac{m}{2m-2}\Pbb(\Ecal|X_1=1,X_2=1,\text{0 flipped to 1})\right]\\
        &= (1 - p^{m-2}q^m)\left[ \frac{m-2}{2m-2}\Pbb(\Ecal|X_1=1,X_2=0) + \frac{m}{2m-2}\Pbb(\Ecal|X_1=1)\right]
    \end{align*}

    Altogether, we obtain a system of three equations with three unknowns. 
    % Denoting shorthand notation $P_{1,10} := \Pbb(\Ecal|X_1=1,X_2=0)$ and $P_{1,11} := \Pbb(\Ecal|X_1=1,X_2=1)$, and substituting $p=q=1/2$:
    % \begin{align*}
    %     P_1 &= \frac{1}{2}P_{1,10} + \frac{1}{2}P_{1,11}\\
    %     P_{1,10} &= \left(\frac{1}{2}\right)^{2n-2} + \frac{1}{2}\left(1 - \left(\frac{1}{2}\right)^{2n-2}\right)(P_1 + P_{1,11})\\
    %     P_{1,11} &= \left(1 - \left(\frac{1}{2}\right)^{2m-2}\right)\left[\frac{m}{2m-2}P_1 + \frac{m-2}{2m-2}P_{1,10}\right]
    % \end{align*}
    Substituting $p=q=1/2$ and solving the linear system of equations for $n = 5, m = 6$ yields $P_{1,10} = 0.7895$, $P_{1,11} = 0.7884$, and $P_1 = 0.7889$. For further details, see~\cite{han21book}.
\end{example}

The discussion of the simpler~\exa{brute_force} and~\exa{brute_force_ctd} segways naturally to more complicated pattern sequences and broader probability distributions of $X_i$, i.e. those that take more than two values. The next theorem provides a result for when there is dependence among the values of $X_i$.

\begin{definition}[Pattern Overlap]
    For a renewal process $X_1, X_2,\cdots$ taking values $x_1, x_2, \cdots$ from a certain probability distribution, we say that a pattern $(x_{(1)}, x_{(2)}, \cdots, x_{(m)})$ has an overlap of size $k < m$ if
    \begin{align*}
        k := \max\{\ell < m | (x_1, \cdots, x_{\ell}) = (x_{m-\ell+1}, \cdots, x_m)\}
    \end{align*}
    That is, $k$ is the largest value such that the first $k$ elements are identical to the last $k$ elements.
    % A pattern $(x_1,\cdots, x_m)$ has overlap of size $k$ if:
    % \begin{align*}
    %     k = \max\{j < m : (x_{m-j+1},\cdots,x_m) = (x_1, \cdots, x_j)\}
    % \end{align*}
    % That is, there exists a size $j$ subsequence of the pattern at the end which matches another size $j$ subsequence of the pattern at the front. 
\end{definition}

\begin{theorem}[Pattern Occurrence for Discrete-State Markov Chain]\label{thm:pattern_occurrence_mc}
    Let $X_1, X_2,\cdots$ denote a sequence of random variables taking values from a discrete-state Markov Chain with transition probability matrix $P := \{P_{ij}\}$ and stationary distribution $\{\pi_k\}$. Let $T$ denote the next time the pattern $(x_1, \cdots, x_m)$ occurs after its first occurrence. 
    % In this problem, we are interested in the expected value of $T$. Within the specific context of the pattern occurrence problem, we refer to a ``renewal'' as the event when the pattern string $(x_1, \cdots, x_m)$ repeats itself. Hence, the interarrival times of such a renewal process are distributed in the same manner as $T$.
    Then:
    \begin{itemize}
        \item When the pattern $(x_1,\cdots,x_m)$ does not contain any overlaps. then the expected time $T$ in between consecutive observations of the pattern is given by
        \begin{align}\label{eq:T_no_overlap_mc}
            \Ebb[T|X_0 = x_0] = \frac{1}{\pi_{x_1}\prod\limits_{i=1}^{m-1} P_{x_i,x_{i+1}}} + \mu(x_0,x_1) - \mu(x_m,x_1)
        \end{align}
        where $\mu(x,y)$ denotes the mean time it takes to reach state $y$ in the Markov chain from state $x$.

        \item When the pattern $(x_1,\cdots,x_m)$ has an overlap $(x_1,\cdots,x_k)$ of size $k$, and $(x_1,\cdots,x_k)$ itself does not contain any overlaps, then the expected time $T$ in between consecutive observations is given by
        \begin{align}\label{eq:T_overlap_mc}
            \Ebb[T|X_0 = x_0] = \frac{1}{\pi_{x_1}\prod\limits_{i=1}^{m-1} P_{x_i,x_{i+1}}} + \frac{1}{\pi_{x_1}\prod\limits_{i=1}^{k-1} P_{x_i,x_{i+1}}} + \mu(x_0,x_1) - \mu(x_m,x_1)
        \end{align}
    \end{itemize}
\end{theorem}

\begin{example}[\thm{pattern_occurrence_mc} for iid Sequence]\label{ex:pattern_occurrence_iid}
    We can specialize the results of~\thm{pattern_occurrence_mc} to the case where the random variables $X_i$ are iid. Let $X_1, X_2,\cdots$ denote a sequence of iid random variables taking on values from a discrete, finite set $\Xcal$. Denote $p_k := \Pbb(X_i= k)$ for all $i$, and let $T$ denote the next time the pattern $(x_1, \cdots, x_m)$ occurs after its first occurrence. Then:
    \begin{itemize}
        \item When the pattern $(x_1,\cdots,x_m)$ does not contain any overlaps. then the expected time $T$ in between consecutive observations of the pattern is given by
        \begin{align}\label{eq:T_no_overlap_iid}
            \Ebb[T] = \frac{1}{\prod\limits_{i=1}^m p_{x_i}}
        \end{align}

        \item When the pattern $(x_1,\cdots,x_m)$ has an overlap $(x_1,\cdots,x_k)$ of size $k$, and $(x_1,\cdots,x_k)$ itself does not contain any overlaps, then the expected time $T$ in between consecutive observations is given by
        \begin{align}\label{eq:T_overlap_iid}
            \Ebb[T] = \frac{1}{\prod\limits_{i=1}^m p_i} + \frac{1}{\prod\limits_{i=1}^{k} p_i}
        \end{align}
    \end{itemize}
\end{example}

Note that
% , when we return to the motivating example and compute 
computing the expected times to between consecutive instances of, respectively, the $A_1$ and $A_2$ used in~\exa{brute_force}, we obtain the exact same formulas~\eqn{motive_success_time} and~\eqn{motive_failure_time}
% , derived using brute-force. More s
Specifically, $A_2$ does not have any overlaps, which is why $\Ebb[T_2]$ can be derived using the formula~\eqn{T_no_overlap_iid}; $A_1$ has overlapping patterns in the form of all length-$2j$ subsequences of alternating $(1,0)$, for $1\leq j\leq n-1$, and so $\Ebb[T_1]$ can be derived using the formula~\eqn{T_overlap_iid}.

\begin{remark}
    Some of the standard references in renewal theory~\cite{ross96book,ross06book,ross10book} also provide some discussion on the pattern-occurrence problem from renewal theory, but often only present the iid case of~\exa{pattern_occurrence_iid}. In~\thm{pattern_occurrence_mc}, we extended this result to include a possible dependence structure for its relevance to the specific fault-tolerance and congestion control case studies we investigate later in~\sec{case_studies}. While a rough sketch of the proof for the iid case~\exa{pattern_occurrence_iid} can be found in~\cite{ross96book}, we provide a more detailed derivation of the formulas in our work~\cite{han21book}. The proof to~\thm{pattern_occurrence_mc} can also be found in~\cite{han21book}.
\end{remark}

% The extended version of the problem, which is a problem of interest in more general settings of renewal processes, looks into computing the probability of observing a specific pattern first among a predetermined set of patterns we wish to observe. We investigate the more general problem after addressing this simpler coin tossing problem.
Furthermore, we can address the question of comparing multiple different patterns at once. Let $\{A_1, \cdots, A_M\}$ for $M\in\Nbb^{+}$ be $M$ different patterns of interest, and let $T_{A_i}$ denote the time it takes to observe pattern $A_i$. Denote $T_{\text{min}} := \min\{T_{A_1}, \cdots, T_{A_M}\}$ and denote $P_i$ to be the probability of observing pattern $A_i$ first at time $T_{\text{min}}$. We further denote $T_{A_i|A_j}$ to be the additional time it takes to observe the pattern $A_i$ after $A_j$ has been observed. We are interested in computing the $P_i$ and $\Ebb[T_{\text{min}}]$. 

We further denote $T_{A_i|A_j}$ to be the additional time it takes to observe sequence $A_i$ given that we've already observed $A_j$, for all $i\neq j$. Since all the patterns $\{A_1, \cdots, A_M\}$ are known, both $\Ebb[T_{A_i}]$ and $\Ebb[T_{A_i|A_j}]$ can be computed using either~\thm{pattern_occurrence_mc} or~\exa{pattern_occurrence_iid}. We can then derive a system of equations with these variables as follows:
\begin{align*}
    \Ebb[T_{A_k}] &= \Ebb[T_{\text{min}}] + \Ebb[T_{A_k}-T_{\text{min}}] = \Ebb[T_{\text{min}}] + \sum\limits_{j\neq k}\Ebb[T_{A_k|A_j}]P_j + 0\cdot P_k
\end{align*}
which, combined with the constraint that $P_M := 1 - \sum_{i<M}P_i$ yields $M$ linear equations with $M$ unknowns $P_1, \cdots, P_{M-1}, \Ebb[T_{\text{min}}]$. For the case where $M=2$, we obtain the explicit formula:
\begin{align}\label{eq:min_time_first_prob}
    P_1 &= \frac{\Ebb[T_{A_2}] + \Ebb[T_{A_1|A_2}] - \Ebb[T_{A_1}]}{\Ebb[T_{A_2|A_1}] + \Ebb[T_{A_1|A_2}]},\qquad \Ebb[T_{\text{min}}] = \Ebb[T_{A_2}] - \Ebb[T_{A_2|A_1}]P_1
\end{align}

In relation to~\exa{brute_force_ctd}, applying~\eqn{min_time_first_prob} to the case where $A_1$ is $n$ consecutive $1$'s and $A_2$ is $m$ consecutive $0$'s yields the desired result~\eqn{motive_prob}. Likewise, applying~\eqn{min_time_first_prob} to the case where $A_1 := (1,0,\cdots, 1,0)$ and $A_2 := (1,\cdots,1,0,\cdots,0)$ with $n=5,m=6$ implies that $\Ebb[T_{A_1|A_2}] = \Ebb[T_{A_1}]$ and $\Ebb[T_{A_2|A_1}] = \Ebb[T_{A_2}]$. With $p=q=1/2$, we obtain the numerical values derived at the end of~\exa{brute_force_ctd}.

\begin{remark}[Alternative Pattern-Occurrence Problem with Random Walks]
    If we are interested in keeping track of longer pattern sequences of impulsive disturbances, the pattern-occurrence problem can also be solved using an alternative approach with random walks. Borrowing influence from common motivational problems such as the Gambler's Ruin problem, the theory of random walks can be used to determine the expected time it takes the system to reach certain boundaries imposed on the system state, upon which the modulation control mechanism takes over. For the sake of keeping this paper focused, we defer discussion of the connections between the pattern-occurrence problem and the theory of random walks to~\cite{han21book}. A comprehensive treatment of random walk theory background needed to understand these connections can also be found in~\cite{han21book}.
\end{remark}

% \subsection{Connection to Random Walk Theory}\label{subsec:random_walks}
% \import{arxiv_sections/}{random_walk_renewals.tex}

%%%%%%%%%%%%%%%%%%%%%%%%%%%%%%%%%%%%%%%%%%%%%%%%%%%%%%%%%%%%%%%%%%%%%%%
\section{The Modulation Control Component}\label{sec:modulation_part}
In this section, we describe the second part of the controller synthesis procedure: the development of the actual control action to be taken when a specific state is first recognized by the first part learning component, especially during the event where a recognized pattern sequence of impulsive jumps causes the system to grow more and more unstable. This requires explicit formulation of the problem as an optimal control problem with corresponding performance objectives, and there is a wide diversity of modulation-type controllers that can be used in place of this second part depending on the kind of application considered. For the purposes of the specific fault-tolerance and congestion control applications that we investigate in the next~\sec{case_studies}, we introduce the \textit{impulse control method}, which is a Hamilton-Jacobi-Bellman-based control for jump-diffusion processes that has been described in several references such as~\cite{oksendal_stoch_ctrl,davis10} and used for financial applications. Roughly speaking, the impulse control methodology forces the state of the system down to some safer, lower-energy state once it exceeds a prespecified limit determined by the first part of the hierarchical controller. This section adapts the general theory of the impulse control framework from~\cite{oksendal_stoch_ctrl,davis10} so that we can directly use it to the specific case examples to be studied in~\sec{case_studies}.

\subsection{Stochastic Processes Review}\label{subsec:stoch_background}
The general \textit{Poisson random measure}, typically denoted $N(dt,dy)$ over the space $[0,t]\times E$ with ``jump space'' $E$, is characterized by the intensity measure Leb$\times\nu$, where Leb denotes the standard Lebesgue measure in time and $\nu(dy)$ is the probability measure on $E$ describing the distribution of jumps. The formal definition of the Poisson random measure can be found in Theorem 2.3.5 in~\cite{applebaum09_book} or Definition 3.1 in~\cite{watson16}, and other standard texts pertaining to Poisson processes.
% While most results in the theory of Poisson processes are written with respect to Poisson random measures, our scope in this paper is specifically on the compound Poisson process. 

A process $L(t)$ is said to be a \textit{L\'{e}vy process} if 1) all paths of $L$ are right-continuous and left-limit (rcll), 2) $\Pbb(L(0) = 0) = 1$, and 3) $L$ has stationary and independent increments~\cite{watson16,jeanblanc07}. This implies that both Gaussian white noise processes and compound Poisson processes are L\'{e}vy processes, and that the affine combination of the two, such as the process described in~\eqn{gen_levy_sde}, is also a L\'{e}vy process. In fact, a well-known result called the \textit{L\'{e}vy-Khintchine Theorem}, stated formally in Theorem 1.6 of~\cite{watson16}, Theorem 2.7 of~\cite{bass09}, or Theorem 1.2.14 of~\cite{applebaum09}, says that L\'{e}vy measures can be represented as weak limits of the convolution of Brownian motion processes and Poisson random measures. This even includes the L\'{e}vy process whose intensity measure has unbounded jumps; one common example is a Gamma process, which has intensity measure on $\Rbb^{+}$ given by $\nu(dy) = ay^{-1}e^{-by}dy$ such that on any finite interval of time, the number of jumps which lies in the interval $(0,1)$ is infinite. We do not consider such types of L\'{e}vy processes in this paper, since they rarely occur in the practical control and engineering applications of our target scope. Throughout this paper, we henceforth use the phrase ``L\'{e}vy noise processes'' specifically to refer only to the class of \textit{bounded-measure} L\'{e}vy processes of the linear combination form in~\eqn{gen_levy_sde}. 

\begin{lemma}[It\^{o}'s Formula for 1D Jump-Diffusion Processes]
    Consider the SDE
    \begin{align}\label{eq:sample_sde}
        dX(t) = f(t)dt + \sigma(t)dW(t) + \int_{\Rbb} \xi(t,z)N(dt,dz)
    \end{align}
    where $X(t) \in \Rbb$. Then for functions $F\in \Ccal^{(1,2)}$, the derivative is given by
    \begin{align}\label{eq:ito_shot}
        dF(t,X(t)) &= \partial_t F(t,X(t))dt + \partial_xF(t,X(t))dX(t) + \frac{1}{2}\partial_x^2 F(t,X(t)) d[X,X](t)\notag\\
        % &= \partial_t F(t,X(t))dt + \partial_xF(t,X(t))\left( f(t)dt + \sigma(t)dW(t) + \int_{\Rbb} \xi(t,z)N(dt,dz)\right)\notag\\
        % &\hskip1cm + \frac{1}{2}\left(\partial_x^2 F(t,X(t)) \sigma^2(t)dt + 2\int_{\Rbb} \left[ F(t,X(t)+z) - F(t,X(t)) - z\partial_xF(t,X(t))\right] N(dt,dz)\right)\notag\\
        &= \partial_t F(t,X(t))dt + \partial_xF(t,X(t))dX^c(t) + \frac{1}{2}\partial_x^2 F(t,X(t)) d[X,X]^c(t)\notag\\
        &\hskip1cm + \int_{\Rbb} \left[ F(t,X(t)+z) - F(t,X(t))\right] N(dt,dz)
    \end{align}
    where $dX^c(t)$ represents the continuous part of the SDE, and $d[X,X]^c(t)$ represents the continuous part of the quadratic variation.
\end{lemma}

\begin{proof}
    We provide a proof of It\^{o}'s formula which is simpler than the version presented in~\cite{protter_book}. Note that by integration-by-parts,
    \begin{align*}
        &d(X(t)Y(t)) = X(t-)dY(t) + Y(t-)dX(t) + d[X,Y](t)\\
        \Longrightarrow \ &[X,Y](t) = X(t)Y(t) - \int_0^t X(s-)dY(s) + Y(s-)dX(s)
    \end{align*}
    where $X(t)$ and $Y(t)$ are two separate stochastic processes and subsequently,
    \begin{align*}
        [X,X](t) = X^2(t) + 2\int_0^t X(s-)dX(s)
    \end{align*}

    Hence
    \begin{align*}
        \Delta[X,X](t) &= \Delta(X^2(t)) + 2X(t-)\Delta X(t)\\
        &= X^2(t) - X^2(t-) + 2X(t-)(X(t) - X(t-)) = \left( \Delta X(t)\right)^2
    \end{align*}

    Recall the second-order Taylor expansion for $F\in\Ccal^2$. We further assume $x \leq y$:
    \begin{align*}
        F(y) - F(x) &= \partial_xF(x)(y-x) + \frac{1}{2}\partial_x^2F(x)(y-x)^2 + R(y,x)
    \end{align*}
    where the remainder term $R(y,x)$ can be written as 
    \begin{align*}
        R(y,x) = \frac{\partial_x^{3}(z)}{3!}(y-x)^{3} \text{ for some } z \in (x,y)
    \end{align*}

    Now let $\Pcal_n := \{ 0 =: T_0^n \leq T_1^n \leq \cdots \leq T_{m_n}^n := t\}$ be a partition of the time interval $[0,t]$. We can then write the difference as the sum
    \begin{align}\label{eq:riemann_sum_ito_proof}
        F(X(t)) - F(x_0) = \sum\limits_{i=0}^{m_n-1} \left[ F(X(T_{i+1}^n)) - F(X(T_{i}^n))\right]
    \end{align}

    We split the analysis into two cases
    \begin{itemize}
        \item \textit{Continuous Case}: When $X(t)$ is the state of a SDE which only has continuous terms, we can use Taylor's formula to expand every term in the sum~\eqn{riemann_sum_ito_proof}:
        \begin{align*}
            \eqn{riemann_sum_ito_proof} &= \sum\limits_{i=0}^{m_n-1} \partial_xF(X(T_{i}^n))(X(T_{i+1}^n) - X(T_{i}^n))\\
            &\hskip1cm + \frac{1}{2}\sum\limits_{i=0}^{m_n-1}\partial_x^2F(X(T_{i}^n))\left(X(T_{i+1}^n) - X(T_{i}^n)\right)^2\\
            &\hskip1cm + \sum\limits_{i=0}^{m_n-1} R(X(T_{i+1}^n), X(T_{i}^n))
        \end{align*}
        and as $n\to\infty$, i.e., as the interval size between partitions decreases to $0$,
        \begin{align*}
            \sum\limits_{i=0}^{m_n-1} \partial_xF(X(T_{i}^n))(X(T_{i+1}^n) - X(T_{i}^n)) &\to \int_0^t \partial_xF(X(s-))dX(s)\\
            \frac{1}{2}\sum\limits_{i=0}^{m_n-1}\partial_x^2F(X(T_{i}^n))\left(X(T_{i+1}^n) - X(T_{i}^n)\right)^2 &\to \frac{1}{2}\int_0^t \partial_x^2F(X(s-))d[X,X](s)
        \end{align*}
        and $R(X(T_{i+1}^n), X(T_{i}^n))\to 0$ since 
        \begin{align*}
            R(X(T_{i+1}^n), X(T_{i}^n)) \propto (X(T_{i+1}^n) - X(T_{i}^n))^3 \text{ and } X(T_{i+1}^n) - X(T_{i}^n)\to 0 \text{ as } n\to\infty
        \end{align*}
        With the terms all combined altogether
        \begin{align}\label{eq:ito_cts}
            F(X(t)) - F(x_0) = \int_0^t \partial_xF(X(s-))dX(s) + \frac{1}{2}\int_0^t \partial_x^2F(X(s-))d[X,X](s)
        \end{align}
        which is indeed the It\^{o} formula when the SDE contains only continuous terms.

        \item \textit{General Semimartingale Case}: We split the intervals of the partition into two pieces: $\Pcal_n^{+}$ is the subset of the subintervals over which the trajectory $X(t)$ contains a jump of size at least $0 < \varepsilon << 1$ and $\Pcal_n^{-} := \Pcal_n / \Pcal_n^{+}$ is the subset of subintervals over which the trajectory $X(t)$ is completely continuous.
        \begin{align*}
            \Pcal_n^{+} := \{ T_{i}^n \ | \ \exists s \in (T_{i}^n, T_{i+1}^n] \ \text{ s.t. } \ \abs{X(s) - X(s-)} > \varepsilon\}
        \end{align*}

        Then we split the sum~\eqn{riemann_sum_ito_proof} into two parts:
        \begin{align*}
            \eqn{riemann_sum_ito_proof} &= \sum\limits_{i\in\Pcal_n^{+}}^{m_n-1} \left[ F(X(T_{i+1}^n)) - F(X(T_{i}^n))\right] + \sum\limits_{i\in\Pcal_n^{-}}^{m_n-1} \left[ F(X(T_{i+1}^n)) - F(X(T_{i}^n))\right]
        \end{align*}

        Note that as $n\to\infty$, the second collective summation term can be reduced to the Continuous Case analyzed previously, with a superscript of $c$ appended to the $dX$ and $d[X,X]$ terms to distinguish the continuous part from the discontinuous part. Furthermore, as $n\to\infty$, the first collective sum converges as follows
        \begin{align*}
            \sum\limits_{i\in\Pcal_n^{+}}^{m_n-1} \left[ F(X(T_{i+1}^n)) - F(X(T_{i}^n))\right] \to \sum_{0\leq s\leq t} \left[ F(X(s)) - F(X(s-))\right]
        \end{align*}
    \end{itemize}
    With all the terms combined together, we obtain the It\^{o} formula described in~\eqn{ito_shot}. This concludes the proof.
\end{proof}

We further define the compensated Poisson random measure as $\tilde{N}(dt,dz) := N(dt,dz) - \nu(dz)dt$, where $\nu(dz)dt := \Ebb\left[ N(dt,dz)\right]$. We often care about the compensated Poisson random measure more than the original measure $N$ because it is mean zero. For jump-diffusions of the form
\begin{align}\label{eq:sample_sde_compensated}
    dX(t) &= f(t)dt + \sigma(t)dW(t) + \int_{\Rbb} \xi(t,z)\tilde{N}(dt,dz)\notag\\
    &= \left(f(t) - \int_{\Rbb}\xi(t,z)\nu(dz)\right) dt + \sigma(t)dW(t) + \int_{\Rbb}\xi(t,z)N(dt,dz)
\end{align}
the It\^{o} formula is written as follows
\begin{align*}
    dF(t,X(t)) 
    % &= \partial_t F(t,X(t))dt + \partial_xF(t,X(t))dX^c(t)\\
    % &\hskip1cm + \frac{1}{2}\partial_x^2 F(t,X(t)) d[X,X]^c(t) + \int_{\Rbb} \left[ F(t,X(t)+z) - F(t,X(t))\right] N(dt,dz)\\
    % &= \partial_t F(t,X(t))dt + \partial_xF(t,X(t))\left(f(t) - \int_{\Rbb}\xi(t,z)\nu(dz)\right)\\
    % &\hskip2cm + \frac{1}{2}\partial_x^2 F(t,X(t)) \sigma^2(t)dt + \int_{\Rbb} \left[ F(t,X(t)+z) - F(t,X(t))\right] (\tilde{N}(dt,dz) + \nu(dz)dt)\\
    &= \left(\partial_t F(t,X(t)) + \partial_xF(t,X(t))f(t) + \frac{1}{2}\partial_x^2 F(t,X(t)) \sigma^2(t)\right)dt\\
    &\hskip1cm + \int_{\Rbb} \left[ F(t,X(t)+z) - F(t,X(t)) - \partial_xF(t,X(t))\xi(t,z)\right]\nu(dz)dt\\
    &\hskip1cm + \partial_xF(t,X(t))\sigma(t)dW(t) + \int_{\Rbb} \left[ F(t,X(t)+z) - F(t,X(t))\right]N(dt,dz)
\end{align*}

% Note the extra term in the integrand of the integral with respect to $N(dt,dz)$. We can obtain an equivalent It\^{o} formula with $\overline{N}(dt,dz)$ instead.

\begin{definition}[Infinitesimal Generator]
    For $F\in\Ccal^{(1,2)}$, the \textit{infinitesimal generator} is defined to be
    \begin{align*}
        \Lcal F(x) = \lim_{t\to 0} \frac{\Ebb_x\left[ F(X(t))\right] - F(x)}{t}
    \end{align*}
    where $\Ebb_x[F(X(t))]$ denotes the expected value of $F(X(t))$ evaluated with initial condition $X(0) = x$.
\end{definition}

From It\^{o}'s formula, we can determine the form of the infinitesimal generator for the SDE~\eqn{sample_sde}. Note that because the integral with respect to the white noise process $W(t)$ and the integral with respect to the compensated Poisson process $\tilde{N}$ are martingales
\begin{align*}
    \Lcal F(x) = \partial_t F(t,X(t)) + \partial_xF(t,X(t))f(t) + \frac{1}{2}\partial_x^2 F(t,X(t)) \sigma^2(t)
\end{align*}

\begin{theorem}[Dynkin's Formula]\label{thm:dynkin}
    For $X(t)$ a trajectory of~\eqn{sample_sde} and $\varphi\in\Ccal^2$, the following equality holds:
    \begin{align*}
        \Ebb_x\left[ \varphi(X(t))\right] = \varphi(x) + \Ebb_x\left[ \int_0^{\tau_S} \Lcal\varphi(X(s))ds\right]
    \end{align*}
\end{theorem}

\subsection{The Impulse Control Method}\label{subsec:impulse_control}
Consider the following L\'{e}vy noise system, assuming the control input enters into the system additively
\begin{align*}
    d\xvect(t) = f(\xvect(t))dt + \sigma(\xvect(t))dW(t) + \int_{\Rbb^{\ell}} \xi(\xvect(t-),z)\tilde{N}(dt,dz) + u(t)dt
\end{align*}
An \textit{impulse control law} is defined to be a sequence of intervention times and corresponding impulse heights $u := (\tau_1, \tau_2, \cdots; z_1, z_2, \cdots)$ such that the closed-loop system state $\xvect_u(t)$ abides by the following dynamics:
\begin{align*}
    \xvect_u(0-) &= y\\
    d\xvect_u(t) &= f(\xvect_u(t))dt + \sigma(\xvect_u(t))dW(t) + \int_{\Rbb^{\ell}} \xi(\xvect_u(t-),z)\tilde{N}(dt,dz) + \sum\limits_j \mathds{1}\{t - \tau_j\}z_j
\end{align*}
% where the closed-loop behavior of the system prior to a jump impulse is abstracted away into the function $\Gamma$.

As with many standard optimal control problems, we seek to find an optimal impulse control law which minimizes a certain cost. To this end, we denote the augmented state $\yvect(t) := (t,\xvect(t))^T$. Define the cost-to-go function $\ell(\yvect)$ and the terminal cost $g(\yvect(t))$. Further denote the cost of making an intervention with impulse $z$ and state $y$ as $K(\yvect,z)$. Then the overall performance objective to minimize is as follows:
\begin{align*}
    J_u(y) = \Ebb_{y}\left[ \int_0^{\tau_S} \ell(\yvect_u(t))dt + g(\yvect_u(\tau_S))\mathds{1}\{\tau_S < \infty\} + \sum\limits_{\tau_j \leq \tau_S} K(\yvect_u(\tau_j), z_j)\right]
\end{align*}
where $\tau_S$ is the (possibly infinite) stopping time of the system and $y$ is the initial state.

As is standard in many Hamilton-Jacobi optimal control approaches, the problem is posed as the following minimization scheme: we want to find $\Phi(y)$ and $u^{*} \in \Ucal$ such that
\begin{align*}
    \Phi(y) := \inf_{u\in\Ucal} J_u(y) = J_{u^{*}}(y)
\end{align*}

A verification theorem provides for a tractable way to iteratively design such a $u^{*}$ with guaranteed optimality. A more general version of the verification theorem with certain continuity, differentiability, and convergence conditions imposed on any candidate value function $\varphi:\Rbb^{+}\times\Rbb^n\to\Rbb^{\geq 0}$ such that $\varphi(y) \geq \Phi(y)$ for all $y$ is presented in~\cite{oksendal_stoch_ctrl,davis10}, but it is too complex for the purposes of our controller synthesis procedure.~\thm{impulse_verification} introduces a simplified version of the verification theorem. For the case studies we investigate in~\sec{case_studies}, any candidate value function that we consider immediately satisfies the continuity, differentiability, and convergence conditions imposed in the general statement from~\cite{oksendal_stoch_ctrl,davis10}; we thus remove explicit mention of these conditions. More importantly,~\thm{impulse_verification} yields a variational inequality that is easier to solve. For simplicity of notation, the verification theorem is presented with the state space taken to be the real line ($\ell = 1$) with the notation $Y(t) := \yvect(t)$ and $Y_u(t) := \yvect_u(t)$, but extension to multiple dimensions is straightforward. 

\begin{theorem}[Verification Theorem]\label{thm:impulse_verification}
    Define
    \begin{align*}
        \Mcal\varphi(y) := \inf_{z\in\Rbb}\left\{ \varphi(y + z) + K(y,z)\right\}
    \end{align*}
    for Borel-measurable, twice continuously-differentiable function $\varphi$.
    \begin{enumerate}
        \item[A.] Suppose
        \begin{enumerate}
            \item $\Lcal\varphi(y) + \ell(y) \leq 0$ for all $y\in D$ where $D := \{ y\in\Rbb | \varphi(y) \geq \Mcal\varphi(y)\}$
            % \item $\varphi(y) \geq \Mcal\varphi(y)$ for all $y \in \Rbb$
            \item $\varphi(Y(t))\to g(Y(\tau_S))\mathds{1}\{\tau_S < \infty\}$ as $t\to\tau_S$
        \end{enumerate}
        Then $\varphi(y) \geq \Phi(y)$ for all $y \in \Rbb$. 
        % Note that the two conditions and the definition of $D$ imply that $\varphi(y) = \Mcal\varphi(y)$ for all $y\not\in D$.

        \item[B.] Put $\tau_0 = 0$ and construct $\hat{u}$ inductively by
        \begin{align*}
            \hat{\tau}_{j+1} &= \min\{\tau_S, \inf\{ t > \hat{\tau}_j | Y_{\hat{u}_j}(t) \not\in D\}\}\\
            \hat{z}_{j+1} &= \hat{\zeta}(Y_{\hat{u}_j}(\hat{\tau}_{j+1}-))
        \end{align*}
        where $\hat{\zeta}(y) \in \text{argmin}_z\Mcal\varphi(y)$ and $Y_{\hat{u}_j}$ is the result of applying $\hat{u}_j := (\tau_1, \cdots, \tau_j; z_1, \cdots, z_j)$ to $Y$.

        Suppose that in addition to A, we have
        \begin{enumerate}
            \item $\Lcal\varphi(y) + \ell(y) = 0$ for all $y\in D$
            \item $\hat{u} \in \Ucal$ and $\{\varphi(Y_{\hat{u}}(t))\}$ is uniformly integrable.
        \end{enumerate}
        Then $\varphi(y) = \Phi(y)$ for all $y\in\Rbb$.
    \end{enumerate}
\end{theorem}

\begin{proof}
    To prove part A, we recall Dynkin's formula from~\thm{dynkin}. Applying this formula to consecutive intervals $[\tau_j, \tau_{j+1})$ which each contain a single impulse jump yields:
    \begin{align*}
        \Ebb_y\left[ \varphi(Y_{u}(\tau_{j+1}-))\right] - \Ebb_y\left[ \varphi(Y_{u}(\tau_j))\right] = \Ebb_y\left[ \int_{\tau_j}^{\tau_{j+1}} \Lcal\varphi(Y_{u}(s))ds\right]
    \end{align*}

    Fix a specific $m\in\Nbb^{+}$. Then summing all equations of the above form from $j = 0$ to $m$ gives
    \begin{align}\label{eq:verification_eq_1}
        -\varphi(y) - \sum\limits_{j=1}^m \Ebb_y\left[\varphi(Y_{u}(\tau_j)) - \varphi(Y_{u}(\tau_j-))\right] + \Ebb_y\left[\varphi(Y_{u}(\tau_{m+1}-))\right] = \Ebb_y\left[ \int_0^{\tau_{m+1}} \Lcal\varphi(Y_u(s))ds\right]
    \end{align}

    By definition of $\Mcal$, we have
    \begin{align*}
        \Mcal\varphi(Y_u(\tau_j-)) \leq \varphi(Y_u(\tau_j)) + K(Y_u(\tau_j-),z_j)
    \end{align*}
    and subtracting both sides by $\varphi(Y_u(\tau_j-))$:
    \begin{align}\label{eq:verification_eq_2}
        \Mcal\varphi(Y_u(\tau_j-)) - \varphi(Y_u(\tau_j-)) \leq \varphi(Y_u(\tau_j)) - \varphi(Y_u(\tau_j-)) + K(Y_u(\tau_j-),z_j)
    \end{align}

    Substituting the left side of~\eqn{verification_eq_2} into~\eqn{verification_eq_1} and multiplying the inequality throughout by $-1$:
    \begin{align}\label{eq:verification_eq_3}
        &\varphi(y) + \sum\limits_{j=1}^m \Ebb_y\left[\Mcal\varphi(Y_u(\tau_j-)) - \varphi(Y_u(\tau_j-)) - K(Y_u(\tau_j-),z_j)\right]\notag\\
        &\hskip1cm - \Ebb_y\left[\varphi(Y_{u}(\tau_{m+1}-))\right] \geq -\Ebb_y\left[ \int_0^{\tau_{m+1}} \Lcal\varphi(Y_u(s))ds\right]
    \end{align}

    Using the first condition, we have 
    \begin{align*}
        &\sum\limits_{j=1}^m \Ebb_y\left[\Mcal\varphi(Y_u(\tau_j-)) - \varphi(Y_u(\tau_j-))\right] \leq 0\\
        &-\Ebb_y\left[ \int_0^{\tau_{m+1}} \Lcal\varphi(Y_u(s))ds\right] \geq \Ebb_y\left[ \int_0^{\tau_{m+1}} \ell(Y_u(s))ds\right]
    \end{align*}
    which transforms~\eqn{verification_eq_3} to
    \begin{align}\label{eq:verification_eq_4}
        \varphi(y) \geq \Ebb_y\left[ \int_0^{\tau_{m+1}} \ell(Y_u(s))ds\right] + \Ebb_y\left[\varphi(Y_{u}(\tau_{m+1}-))\right] + \sum\limits_{j=1}^m \Ebb_y\left[ K(Y_u(\tau_j-),z_j)\right]
    \end{align}

    Taking $m\to\infty$, i.e. $t\to\tau_S$, and applying it to~\eqn{verification_eq_4} gives
    \begin{align*}
        \varphi(y) \geq \Ebb_y\left[ \int_0^{\tau_{S}} \ell(Y_u(s))ds + g(\tau_{S})\mathds{1}\{\tau_S < \infty\} + \sum\limits_{\tau_j < \tau_S} K(Y_u(\tau_j-),z_j)\right] =: \Phi(y)
    \end{align*}
    and this proves part A of the theorem. To prove part B, we can simply repeat the proof of part A while replacing every inequality with an equality, which comes from the conditions of part B in the theorem's hypothesis.
\end{proof}

Overall, the optimal value function $\Phi(y)$ should satisfy the following variational inequality for all $y\in\Rbb$.
\begin{align*}
    \max \{ \Lcal\varphi(y) + \ell(y), \varphi(y) - \Mcal\varphi(y)\} = 0
\end{align*}

Further illustrative examples demonstrating how to apply the impulse control method for certain jump-diffusion systems is presented in our textbook preprint~\cite{han21book}. For additional background on the connection between the impulse control approach and other standard Hamilton-Jacobi-Bellman optimal control methods, we refer the interested reader to~\cite{han21book}.

\section{Application Case Studies}\label{sec:case_studies}
\subsection{Incremental Stability for Jump-Linear System}\label{subsec:1D_linear}

Now that we have established the relevant theory for the proposed two-part controller, we demonstrate its usage using shot-noise-modified extensions of two simple scalar systems which are commonly studied in the field of stochastic processes. In both examples, stability bounds are explicitly computed using the framework discussed in our first part of this paper~\cite{han21tac}, then used as the certified safety bound which determines the threshold at which the system transitions between two independent controllers. This hierarchical layering of controllers is motivated by spacecraft control applications~\cite{franchi18,wander13,kolcio16}, which manipulates the thrusters by switching between different flight control systems depending on factors such as how far away the system is from a desired trajecory or how close the system is to an impending meteorite strike. When the system remains within the safety region, a lower-level controller (e.g., a reference-tracking controller) drives the trajectory. Moreover, previous incremental stability results for deterministic~\cite{lohmiller98} and Gaussian white noise-perturbed systems~\cite{dani15} show that such systems are guaranteed to remain within the safety region. Hence, for our systems of consideration,~\eqn{gen_shot_sde} and~\eqn{gen_levy_sde}, the shot noise component of the noise process is arguably the main cause of destabilization in the system. Thus, when the system no longer remains within the safety region, the two-part controller is used to first identify the sequence of impulsive disturbances which caused the destabilization, then steer the trajectory back to within the safety region.
% One intuitive example of the fault-tolerance problem is the Tacoma Narrows Bridge which collapsed in 1940. While the bridge was designed to withstand winds as strong as hurricanes, it collapsed as a result of a steady accumulation of small breezes.

% We compute the contraction rate and the steady-state error bound for this example using the theoretical stability framework discussed in~\cite{han20tac}. We recall that incremental stability for deterministic systems has been established as a concept of convergence between different solution trajectories with different initial conditions~\cite{lohmiller98,aminzare14}. 
% We are interested in the problem where given a specific error bound $\varepsilon \geq 0$ and compact support length $c_0$, what is the largest possible $\lambda$ and $\eta$ such that the error bound is satisfied, i.e.,~\eqn{single_jump_bound}$\leq \varepsilon$?
The first scalar system of our study is as follows. We assume the control input enters into the system additively. It can be viewed as the \textit{Ornstein-Uhlenbeck process}~\cite{oksendal_book} augmented with shot noise instead of the usual white noise:
\begin{align}\label{eq:1D_shot_system}
    dx(t) = ax(t)dt + u(t)dt + \xi\int_{\Rbb} z N(dt,dz)
\end{align}
where $a \neq 0$, $\xi > 0$ are constants, and $N(dt,dz)$ is a 1D Poisson random measure with rate $\lambda > 0$ and jump height distribution as a Bernoulli random variable which takes value $\eta > 0$ with probability $p$, and $-\eta$ with probability $q := 1 - p$. As motivated above, we distinguish the control law into two parts
\begin{align*}
    u(t) := \mathds{1}\{x(t) \in \Scal\}u_{\ell}(t) + \mathds{1}\{x(t) \not\in \Scal\}u_{tp}(t)
\end{align*}
where $\Scal \subset \Rbb$ defines the safety region computed from the bound of~\cite{han21tac}, $u_{\ell}(t)$ describes the lower-level controller, and $u_{tp}(t)$ describes the two-part controller. For the purposes of keeping the discussion relevant to the paper, we focus on the degisn and impact of $u_{tp}(t)$ on the system and assume that $u_{\ell}(t)$ is already given. One possible choice is a reference-tracking controller designed to track some reference trajectory $x_r(t)$:
\begin{align}
    u_{\ell,r}(t) = \dot{x}_r(t) - ax_r(t), \qquad u_{\ell}(t) = u_{\ell,r}(t) - k(x(t) - x_r(t))
\end{align}
and gain $k > \abs{a}$. 
% For a less restrictive example, we can impose the system to be controllable, then append to the lower-level controller a state-feedback law of the form $u(t) = -kx(t)$ such that $a - k < 0$, but we choose to look at an open-loop stable system with $a < 0$ from the start for its notational simplicity.

We now explicitly compute the form of the safety region $\Scal$. The contraction metric $M(t,x)$ is chosen to be the identity $1$, meaning $\overline{m} = \underline{m} = 1$, and we have contraction rate $\alpha := \abs{a}$. We substitute only the lower-level controller $u_{\ell}(t)$ into~\eqn{1D_shot_system} for the computation of the stability bound, with the interpretation being that we want to determine the local extent of which the lower-level controller is able to handle the system.
\begin{align*}
    x(t) &= x_0e^{(a-k)t} + \int_0^t \left(u_r(t) + kx_r(t)\right)e^{(a-k)(t-s)}ds + \int_0^t\int_{\Rbb} \xi ze^{(a-k)(t-s)}N(ds,dz)
\end{align*}
A trajectory of the noiseless system is given by
\begin{align*}
    y(t) &= y_0e^{(a-k)t} + \int_0^t \left(u_r(t) + kx_r(t)\right)e^{(a-k)(t-s)}ds
\end{align*}
% where we use  and $T_i$ is the time of the $i$th jump in the Poisson process $n(t)$. 
Taking the mean-squared difference between $x(t)$ and $y(t)$ yields:
\begin{align}\label{eq:single_jump_bound}
    &\Ebb\left[\abs{y(t) - x(t)}^2\right] \leq \Ebb\left[\abs{y_0 - x_0}^2\right]e^{2(a-k)t} + \kappa_s(2(a-k),t)
\end{align}
with steady-state error bound
\begin{align}\label{eq:1Dlqr_kappa_unsimp}
    \kappa_s(2(a-k),t) &= 2e^{(a-k)t}\Ebb\left[\abs{y_0 - x_0}\right]\Ebb\left[\abs*{\int_0^t\int_{\Rbb}\xi z e^{(a-k)(t-s)}N(ds,dz)}\right]\notag\\
    &\hskip1cm + \Ebb\left[\abs*{\int_0^t\int_{\Rbb} \xi z e^{(a-k)(t-s)} N(ds,dz)}^2\right]\notag\\
    &\leq 2\xi e^{(a-k)t}c_0\Ebb\left[\int_0^t\int_{\Rbb}\abs*{z e^{(a-k)(t-s)}}N(ds,dz)\right]\notag\\
    &\hskip1cm + \xi^2\Ebb\left[\int_0^t\int_{\Rbb} \abs*{z e^{(a-k)(t-s)}}^2 N(ds,dz)\right]
    % \notag\\
    % &\hskip3cm + \eta^2\sum\limits_{k=0}^{\infty}\Ebb\left[ \left(\sum\limits_{i=1}^{k} e^{a(t-T_i)}\right)^2\bigg| n(t) = k\right]\Pbb(n(t) = k)
\end{align}
where $c_0 > 0$ is the length of the compact support of the probability distribution where the initial conditions $x_0, y_0$ are drawn from, and the second inequality follows from the Cauchy-Schwarz Inequality. Note that the Poisson integrals in the two terms can be evaluated using Campbell's formula (presented in Section 3.2 of~\cite{kingman_book} and Proposition 2.7 of~\cite{lastpenrose17}) and the definition of the Poisson integral from Section 2.3.2 of~\cite{applebaum09_book}:
\begin{align*}
    \Ebb\left[\int_0^t\int_{\Rbb}\abs*{z e^{(a-k)(t-s)}}N(ds,dz)\right] &= \int_0^te^{(a-k)(t-s)}\int_{\Rbb}\abs{z}\nu(dz)ds = \frac{\eta}{a-k}(e^{(a-k)t} - 1)\\
    \Ebb\left[\int_0^t\int_{\Rbb} \abs*{z e^{(a-k)(t-s)}}^2 N(ds,dz)\right] &= \int_0^te^{(a-k)(t-s)}\int_{\Rbb} z^2 \nu(dz)ds = \frac{\eta^2}{2(a-k)}(e^{2(a-k)t} - 1)
\end{align*}
and substituting back into~\eqn{1Dlqr_kappa_unsimp} yields:
\begin{align}\label{eq:1Dlqr_kappa}
    \eqn{1Dlqr_kappa_unsimp} &\leq \frac{2\xi\eta c_0}{a-k}(e^{2(a-k)t} - e^{(a-k)t}) + \frac{\xi^2\eta^2}{2(a-k)}(e^{2(a-k)t} - 1)
\end{align}

\begin{figure}
    \begin{center}
    \includegraphics[width=0.6\columnwidth]{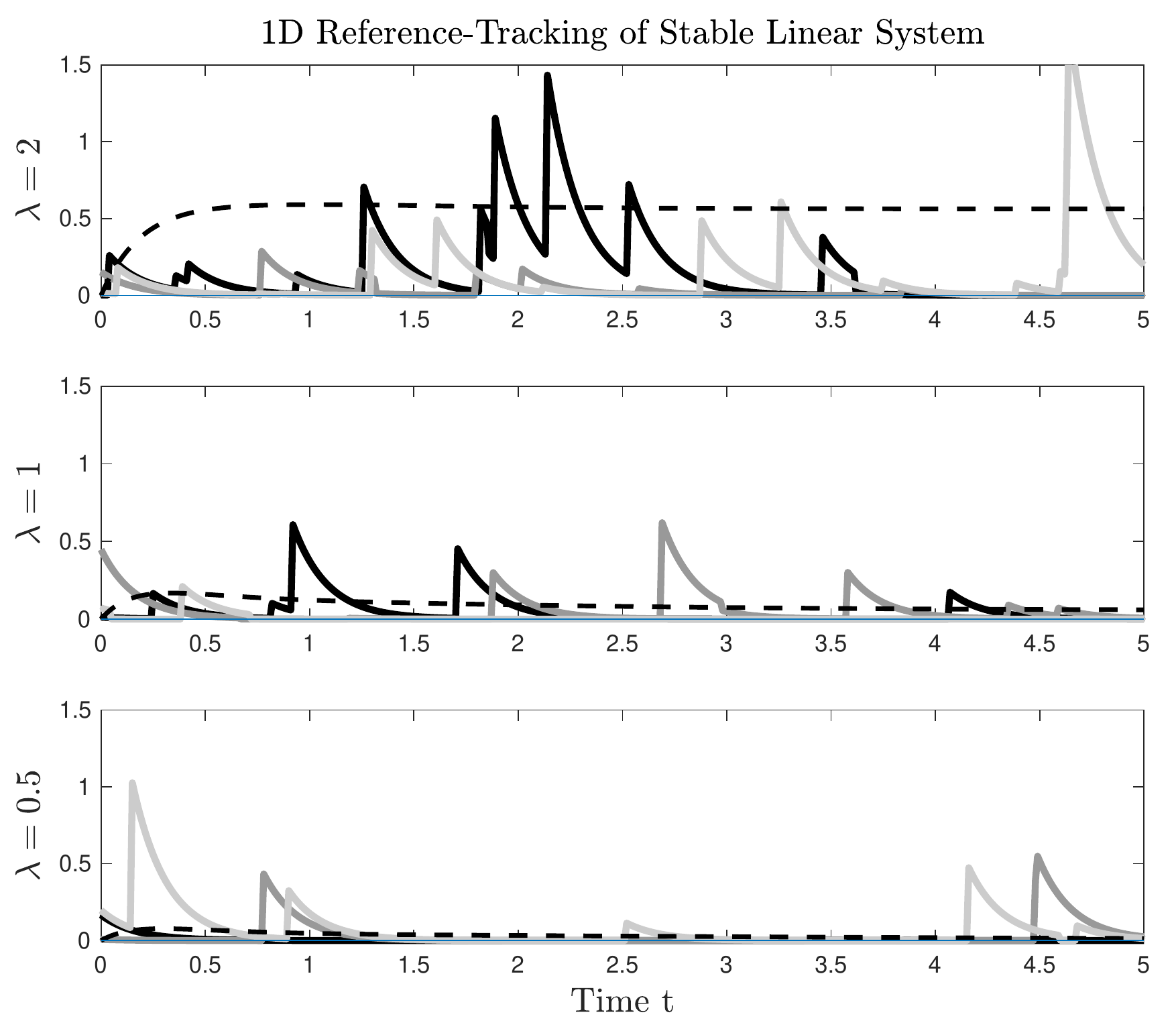}
    \caption{Three sample absolute MSE differences (in varying shades of gray) between the trajectory of a stable linear system and the trajectory of the system with additive shot noise for varying values of $\lambda$. For each subplot corresponding to $\lambda$, the black-dashed line shows the outline of the envelope captured by the theoretical error bound derived as in~\eqn{single_jump_bound}.}
    \label{fig:1D_lqr_vs_lambda}
    \end{center}
\end{figure}

We simulate~\eqn{1D_shot_system} using the state-feedback lower-level controller to track the zero reference line (i.e., perform disturbance rejection), and visualize our results in~\fig{1D_lqr_vs_lambda} for three values of $\lambda \in \{2,1,0.5\}$. For each $\lambda$ value, we simulate three sample trajectories (in three different shades of gray) representing the mean-squared-difference $\abs{y(t) - x(t)}^2$. The initial conditions $x_0$ and $y_0$ are sampled uniformly in the range $[0,2]$. For each subfigure, plotted in black-dashed line with their respective colors is the upper-bound of the envelope captured by the theoretical bound of $\kappa_s(-2(a-k),t)$ provided by~\eqn{1Dlqr_kappa}. Notice that it is entirely feasible for the shot noise perturbations to bring each trajectory outside of the safety envelope, upon which we allow the two-part controller to take over and steer the system back to within the envelope. Note that we achieve global exponential convergence for the system because of its simplified linear nature, so the lower-level controller manages to stabilize the system even while it is outside of the safety envelope by itself. However, we can still demonstrate the procedure of using the learning component of the two-part controller for determining re-ocurring jump patterns on this system; in our second example, we add Gaussian white noise to the dynamics so that the necessity of the two-part controller is more apparent.
% That being said, the simulations demonstrate the conservativeness of $\kappa_s$ bound from~\thm{shot_contract}. In this simple case, even trajectories corresponding to $\lambda = 2$ converge to within a very small distance about zero. Additionally, note in particular, that the steady-state error bound $\kappa_s(-2a,t)$ is proportional to $\eta$ and, as evidenced in~\fig{1D_lqr_vs_lambda}, a larger $\lambda$ corresponds to a larger error bound. These confirm the intuition of~\remk{tradeoff}, that both larger jumps and shorter interarrival times (i.e., larger $\lambda$) correspond to a larger theoretical steady-state error bound. % In order to ensure sufficient convergence of the system trajectory to the reference, the bound relies on longer interarrival times $T_i - T_{i-1}$, which correspond to smaller intensity $\lambda$.

We assume that a given specific pattern sequence of jump sizes leads to instability in the system. In this simple setup, the learning component approximates variations in the interarrival time between consecutive jumps using various zero paddings. What is the number of jumps $T$ after which the exact pattern sequence of jumps will occur again? For concreteness, we focus on the specific pattern of $(1,2,1)$ with $M = 9$. Then
\begin{align}\label{eq:increm_stab_pattern_single}
    \Ebb[T] = \frac{1}{p_1^2p_2} + \frac{1}{p_1}
\end{align}

By symmetry of the problem, negative valued jumps should also be considered, i.e. $(-1,-2,-1)$. In this case, we compute $\Ebb[T]$ by conditioning using the same methodology described in Section~\ref{subsec:pattern_occurrence}. Denote $A:=(1,2,1)$ and $B:=(-1,-2,-1)$ and use the notation $T[(*)] := \Ebb[T(*)]$ and $T[\text{min}] := \Ebb[T_{\text{min}}]$ for simplicity. Similar to~\eqn{increm_stab_pattern_single}, we get:
\begin{align*}
    T[A] = \frac{1}{p_1^2p_2} + \frac{1}{p_1}, \ \ T[B] = \frac{1}{p_{-1}^2p_{-2}} + \frac{1}{p_{-1}}
\end{align*}
and because there is no overlap between $A$ and $B$, the conditional quantities are exactly the same as the original quantities: $T[B|A] = T[B]$ and $T[A|B] = T[A]$.

Denote $P_A$ to be the probability that the string sequence $A$ occurs before $B$. Then the renewal time in between consecutive unstable jump sequences is given by $T_{\text{min}} := \min\{ T_A, T_B\}$, which can be obtained according to Section~\ref{subsec:pattern_occurrence}.

Now, further note that the order of the jump sizes also does not matter, since it is the cumulative sum of the jump sizes which impacts the system trajectory (again, assuming the interarrival times between consecutive jumps are not accounted for). Hence, all three combinations for $(1,2,1)$ and $(-1,-2,-1)$ should be taken into consideration. For the moment, we will only consider the permutations of $(1,2,1)$. Define $A:=(1,2,1)$, $B:=(1,1,2)$, and $C:=(2,1,1)$. $T_{\text{min}}:=\min\{T(A), T(B), T(C)\}$. Computing the expected renewal time $T[\text{min}]$ only requires a slight extension to the procedure of Section~\ref{subsec:pattern_occurrence}, but the idea is the same.

Note that 
\begin{align*}
    T[A] = \frac{1}{p_1^2p_2} + \frac{1}{p_1}, \qquad T[B] = T[C] = \frac{1}{p_1^2p_2}
\end{align*}
and
\begin{align*}
    T[A|B] = T[A], &\qquad T[A|C] = T[A] - T[(1)] = \frac{1}{p_1^2p_2}\\
    T[B|A] = T[B] - T[(1)] = \frac{1}{p_1^2p_2} - \frac{1}{p_1}, &\qquad T[B|C] = T[B] - T[(1,1)] = \frac{1}{p_1^2p_2} - \frac{1}{p_1^2} - \frac{1}{p_1}\\
    T[C|A] = T[C], &\qquad T[C|B] = T[C] - T[(2)] = - \frac{1}{p_1^2p_2} - \frac{1}{p_2}
\end{align*}

Now denote $P_A$ to be the probability that $A$ is the first to occur among all three patterns. Likewise, define $P_B$ and $P_C = 1 - P_A - P_B$. We can write a system of three equations with three unknowns
\begin{align*}
    T[\text{min}] = T[A] + T[\text{min}] - T[A] = T[A] + T[A|B]P_B + T[A|C]P_C\\
    T[\text{min}] = T[B] + T[\text{min}] - T[B] = T[B] + T[B|A]P_A + T[B|C]P_C\\
    T[\text{min}] = T[C] + T[\text{min}] - T[C] = T[C] + T[C|A]P_A + T[C|B]P_B
\end{align*}
% Under the assumption that the probability distribution is uniform over all the values, i.e. $p_i = 1/18$, we obtain the following values:
% \begin{align*}
%     P_A = \frac{5035}{14283} = 0.3525, \ \ P_B = \frac{33055}{104742} = 0.3156, \ \ P_C = \frac{9481}{28566} = 0.3319
% \end{align*}

% Note that a natural next step of the problem is the case where we do not care about the length of the jump sequence pattern as long as the cumulative jump size that results from it does not exceed a specific value. For instance, for jump values $\xi(t)$ restricted to $\{1,2\}$ only, a total jump size of $4$ is achieved by any of the following sequences: $(1,1,1,1)$, $(2,1,1)$ and its combinations, or $(2,2)$.

Now we consider the following L\'{e}vy noise system, which can be viewed as a L\'{e}vy noise extension to the well-known \textit{Geometric SDE}~\cite{oksendal_book}. We demonstrate how to design the control action synthesis part of the two-part controller. 
\begin{align}\label{eq:1D_levy_system}
    dx(t) = ax(t)dt + u(t)dt + \sigma x(t) dW(t) + \xi x(t)\int_{\Rbb} z N(dt,dz)
\end{align}
where $a\neq 0, \sigma > 0, \xi > 0$ are constants, and the jump size $z$ is distributed uniformly over the interval $[-2,-1]\cup[1,2]$. We invoke the impulse control methodology described in Section~\ref{subsec:impulse_control} to determine how to steer the trajectory of the system back towards the bounded error ball after the instability pattern has occurred.
% dependent on the state of the system as follows
% \begin{align*}
%     \xi(x,z) = \begin{cases}
%         +3 &\text{ if } x \geq \varepsilon\\ 0 &\text{ if } \abs{x} < \varepsilon\\ -3 &\text{ if } x \leq -\varepsilon
%     \end{cases}
% \end{align*}
% The control action part of the reference tracker can be designed according to a variation of an example which was presented in~\cite{oksendal_stoch_ctrl}. 
Suppose that we seek to minimize the performance objective designed with the cost-to-go and impulse cost functions chosen as
\begin{align*}
    K(y,z) = e^{-t}5\abs{z}, \qquad \ell(y) = e^{-t}x^2
\end{align*}
We have $y := (t,x)$ to be the state of the augmented system. Consider a value function of the form $\varphi(s,x) = e^{-s}\psi(x)$. Due to the symmetry of the problem, we choose region 
\begin{align*}
    D := \{x\in\Rbb | \psi(x) < \Mcal\psi(x)\} = (-b, b)
\end{align*}
for some $b > 0$ to be determined.
%For the time being, we will assume $b$ is known.

The first condition of~\thm{impulse_verification} states that the following expression needs to be zero for $x \in [-b, b]$ in order for $\varphi$ to be optimal:
\begin{align*}
    \Lcal\varphi(t,x) + \ell(t,x) = e^{-t}\left( -\psi(x) + ax\psi'(x) + \frac{1}{2}\sigma^2x^2\psi''(x) + \int_{\Rbb}\left[ \psi(x + z) - \psi(x)\right]\nu(dz) + x^2\right)
\end{align*}
where $\Lcal\varphi$, i.e. $d\psi$, was computed using It\^{o}'s formula. $\psi$ must then satisfy the following differential equation:
\begin{align*}
    -\psi(x) + ax\psi'(x) + \frac{1}{2}\sigma^2x^2\psi''(x) + \int_{\Rbb}\left[ \psi(x + z) - \psi(x)\right]\nu(dz) = -x^2
\end{align*}
Again, we can solve this nonhomogeneous ordinary differential equation as a sum of homogeneous and nonhomogeneous parts. We guess the exponent of the homogeneous solution $\psi^h(x) := x^r$ through the following auxiliary equation:
\begin{align}\label{eq:1d_impulse_eq1}
    &-1 + ar + \frac{1}{2}\sigma^2r(r-1) + \int_{\Rbb}\left[ (1 + z)^r - 1\right]\nu(dz) = 0
\end{align}

Note that
\begin{align*}
    \int_{\Rbb}\left[ (1 + z)^r - 1\right]\nu(dz) &= \frac{1}{2}\left[\int_{-2}^{-1} \left[ (1 + z)^r - 1\right]dz + \int_{1}^{2} \left[ (1 + z)^r - 1\right]dz\right] = \frac{1}{r+1}\left( (-1)^r + 3^{r+1} - 2^{r+1}\right) - 2
\end{align*}
and so~\eqn{1d_impulse_eq1} reduces to
\begin{align*}
    &-3 + ar + \frac{1}{2}\sigma^2r(r-1) + \frac{1}{r+1}\left( (-1)^r + 3^{r+1} - 2^{r+1}\right) = 0
\end{align*}
% \blue{roots of $r$ come out to be imaginary??

Solving this auxiliary equation yields one positive root $r_{+}$ and one negative root $r_{-}$. Furthermore, the partial solution $\psi^p(x)$ corresponding to the nonhomogeneous part of the equation can be assumed to take on a form of a second-order polynomial $A_2x^2 + A_1x + A_0$. Substitute $\psi(x) = \psi^h(x) + \psi^p(x)$ into the original equation, then match the coefficients to determine the $A_i$. Overall, when $\abs{x} \leq b$, the complete solution turns out to be
\begin{align}\label{eq:symm_jumps_first_half}
    \psi(x) = \psi_1(x) := C_{+}e^{r_{+}x} + C_{-}e^{r_{-}x} + x^2 + 1
\end{align}

Since we have symmetry of the problem in the sense that we desire to push the trajectory within an equidistant tube surrounding the zero line, we choose specifically $C_{+} = C_{-} = -C$, where $C > 0$, so that 
\begin{align*}
    \psi_1(x) = C(e^{r_{+}x} + e^{r_{-}x}) + x^2 + 1
\end{align*}    

When $\abs{x} > b$, we look at the second condition from~\thm{impulse_verification}. First, the intervention operator is written as:
\begin{align}\label{eq:symm_jumps_intervention}
    \Mcal\psi(x) = \inf_{z\in\Rbb} \{ \psi(x+z) + 5\abs{z}\}
\end{align}
which implies that the minimizing value of $z$ satisfies
\begin{align*}
    \begin{cases} \psi'(x+z) = -5 &\text{ if } z > 0\\ \psi'(x+z) = 5 &\text{ if } z < 0 \end{cases}
\end{align*}
and so we want to find $a > 0$ such that $-b < -a < 0 < a < b$. Essentially, the impulse control law is designed to push the trajectory up ($z > 0$) to $-a$ if it goes beneath level $-b$ and push the trajectory down ($z < 0$) to $a$ if it goes above level $+b$. This corresponds to the first-order conditions $\psi'(a) = -5$ and $\psi'(-a) = 5$.

With level $a$ assigned, we explicitly choose the form of 
\begin{align}\label{eq:symm_jumps_second_half}
    \psi_2(x) := \begin{cases} 
        \psi_1(a) + 5(x-a), &\text{ if } x > b\\
        \psi_1(-a) - 5(x+a), &\text{ if } x < -b 
    \end{cases}
\end{align}

Thus, in combination, the form of $\psi$ can be written as follows:
\begin{align}\label{eq:symm_jumps_solution}
    \psi(x) = \begin{cases}
        C(e^{r_{+}x} + e^{r_{-}x}) + x^2 + 1 &\text{ if } \abs{x} \leq b\\
        \psi_1(a) + 5(x-a), &\text{ if } x > b\\
        \psi_1(-a) - 5(x+a), &\text{ if } x < -b 
    \end{cases}
\end{align}

\begin{figure}
    \begin{center}
        \includegraphics[width=0.6\columnwidth]{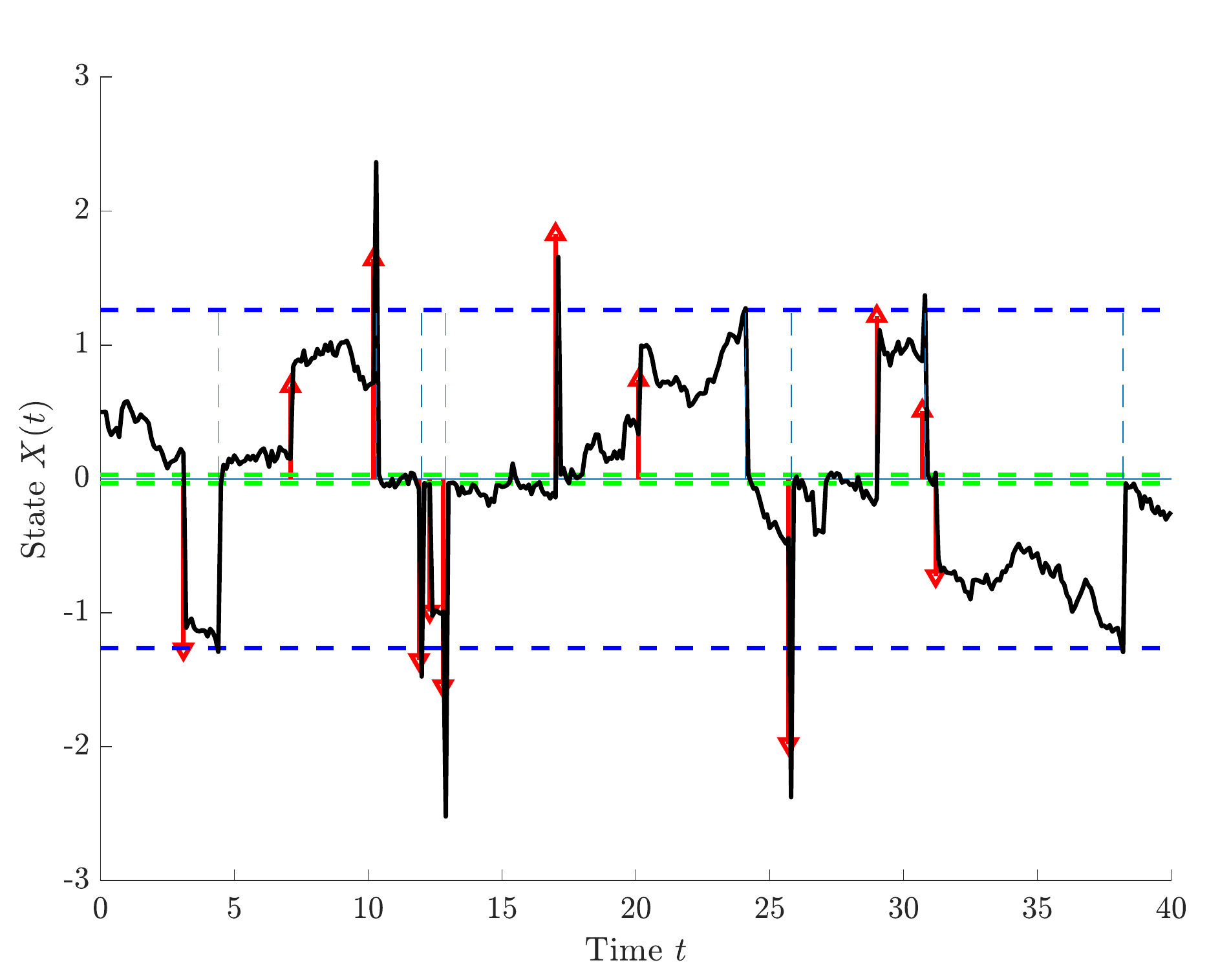}
    \end{center}
    \caption{A sample closed-loop trajectory of the L\'{e}vy noise system with the designed impulse control applied.}
    \label{fig:1D_fault_tol_trajectory}
\end{figure}

Now, we can use the first-order conditions and the continuity and differentiability conditions to solve for $a$ and $C$.
\begin{enumerate}
    \item Continuity at $x=b$ implies we need $\psi(b-) = \psi(b+)$:
    \begin{align*}
        &\psi_1(b) = \psi_1(a) + 5(b-a)\\
        \Longrightarrow \ &C(e^{r_{+}b} + e^{r_{-}b}) + b^2 + 1 = C(e^{r_{+}a} + e^{r_{-}a}) + a^2 + 1 + 5(b-a)
    \end{align*}

    \item Differentiability condition at $x=b$ implies we need $\psi_1'(b) = \psi_2'(b+)$:
    \begin{align*}
        Cr_{+}e^{r_{+}b} + Cr_{-}e^{r_{-}b} + 2b = 5
    \end{align*}

    \item We also have the required condition that $\psi'(a) = \psi_1'(a) = -5$ (and the opposite condition does not need to be mentioned due to symmetry):
    \begin{align*}
        Cr_{+}e^{r_{+}a} + Cr_{-}e^{r_{-}a} + 2a = -5
    \end{align*}
\end{enumerate}

See~\fig{1D_fault_tol_trajectory} for a visualization of a sample trajectory for the system~\eqn{1D_levy_system} with the impulse control law applied. Intuitively, it is easy to see that the impulse control method is optimal because it instantaneously pushes the state of the system down to a safer, lower-energy state. Furthermore, the nature of the control law is that this only happens once the trajectory exceeds a prespecified limit. We remark that the behavior of the trajectory illustrated in~\fig{1D_fault_tol_trajectory} shows the ideal system response under no physical constraints on the system's components, e.g. its actuators. This is similar to the well-known bang-bang control action for spacecraft applications; while theoretically optimal, they are not always practical to implement. Similarly, in this setting, it may be difficult or most practical systems to perform a maneuver as sharp as an impulse control. However, we emphasize that the impulse control mechanism can be replaced with a synthesis procedure that incorporates physical constraints. The hierarchical nature of the two-part controller still remains the same, hence, the benefits arising from storing in memory control laws for future occurrences of destabilizing patterns which have been observed before are still present.

% In order to account for cases of varying jump sequences lengths, we make use of some background from random walk analysis. The following setup discusses a similar problem.
% 7.13 PLACED INTO random_walk_renewals.tex

\subsection{Controlling Vehicle Traffic at an Intersection}\label{subsec:vehicle_traffic}
In this section, we demonstrate application of the two-part controller synthesis procedure on congestion control problems for discrete-valued quantities, such as the number of vehicles present in each lane of a stretch of road, or the number of packets in each flow that is queued up at an internet router. Particularly for congestion control problems, the discrete value of the state allows the learning component of the two-part procedure to identify recurring patterns much more accurately. For that reason, the pattern-occurrence problem is arguably better suited to congestion-type applications moreso than the scalar fault-tolerance control problem presented in the previous section. Another key distinction for how the two-part procedure treats congestion control-type problems is that the control action is applied as soon as possible in an effort to drive the level of congestion close to zero as quickly as possible, as opposed to the fault-tolerance control problems considered previously, which waits until an appropriate accumulation of impulsive disturbances has driven the trajectory beyond a nonzero bound. 

For concreteness, we demonstrate this application to the specific congestion-type problem of controlling vehicle traffic across a single intersection by adjusting light signal patterns, and we look into two specific types of intersection scenarios: 
% 1) the two-way intersection, 
1) the four-way intersection with one lane per road, and 2) the four-way intersection with one lane per road assuming the lanes on the opposite side of the intersection can be congested, and 3) the four-way intersection with two lanes per road, one for forward and right-turning traffic and one for left-turning traffic. For each of the scenarios, we look into how the learning component is designed by using the pattern-occurrence problem to compute the expected time that elapses between two consecutive instances of the same intersection state. First, we establish the general setting and the relevant simplifying assumptions for the problem.

Consider a single four-way intersection for oncoming vehicles. The four ways of each intersection are $E$ for East, $N$ for North, $W$ for West, and $S$ for South. New platoons of vehicles can enter the intersection from any of the four directions according to compound Poisson processes of possibly different intensities, where the jump sizes indicate the number of vehicles entering the lane. For the dynamics, the state we keep track of is the queue length (i.e., number of vehicles) at each of the lanes, represented by the vector $\xvect\in\Rbb^n$. Furthermore, each component $x_\ell \in [0,x_{\text{max}}]$, where $\ell\in\Nbb^{+}$ denotes the ID of the lane in the intersection, and $x_{\text{max}} \in \Zbb^{+}$ denotes the maximum number of cars allowed to enter a single lane within one timestep. For concreteness, we take $x_{\text{max}} = 5$. One simplification we make is the discretization of time and the consequent representation of vehicle interarrival times as Geometric random variables instead of Exponential ones. This allows us to model the dynamics of the number of vehicles per lane of the intersection as a Markov chain. We denote $p_k$ to be the probability that $k\in [1,x_{\text{max}}]$ vehicles enter into lane $i$ within one timestep.
% take, for concreteness, $9/10$ to be the probability of no vehicles entering the $i$th lane at a certain timestep, and and equal probability $1/50$ to be the probability that some number $k \in [1,x_{\text{max}}]$ of vehicles enter into lane $i$ at a certain timestep.

% \begin{assumption}[Particle Vehicles]\label{assum:particle_vehicle}
%     We assume each vehicle has no length and can be treated effectively as a particle. This is to disregard the distance displacement from the intersection as a result of being behind a long queue of vehicles that are stopped there.
% \end{assumption}

The intersection is locally controlled by traffic lights, one per lane, and each take on binary values depending on whether the light is green $(1)$ or not $(0)$. Suppose that the control law signal is such that the probability of the light being green at a certain time instance can be represented by a probability $0 < p_G < 1$ and let $p_R := 1 - p_G$ be the probability that the light is red. This probability is what the second part of the two-part controller adjusts in order to optimize traffic flow, and we discuss this aspect in more detail later on. For simplicity, we assume that all vehicles travel at the same speed, and thus at a single timestep when the light is green, a constant number $M < x_{\text{max}}$ of vehicles can pass. 

The learning component seeks to solve the following problem. Suppose that at time $t$, the state of the intersection is given by $\xvect_0\in\Rbb^n$. We are interested in computing the expected number of timesteps after $t$ in which the state $\xvect_0$ occurs again. This is precisely the pattern-occurrence problem that we've studied in Section~\ref{subsec:pattern_occurrence}; in this setup, the renewals denote the event that the state of the intersection is $\xvect_0$. 
We are interested in this quantity for two complementary reasons. On one hand, it saves computational resources to have the system memorize the optimal traffic signal patterns for the first occurrence of $\xvect_0$, then apply it for every future occurrence of $\xvect_0$. On the other hand, if the system has limited memory storage and the time between consecutive occurrences of $\xvect_0$ is too long, then the optimal control sequence may no longer be retained in memory, prompting recomputation from scratch anyway.
% Compute length-8 state patterns between consecutive renewals. Expected time between consecutive renewals – usual Ross problem. After that, relate back to what was said in the “Current Problems” about how there needs to be sufficient time in between consecutive instability patterns in order for the system to be stable, but at the same time, it shouldn’t take too long because otherwise we won’t be able to use the learning advantage of not needing to recompute the control law.
We now address the pattern-occurrence problem for each of hte three cases described above.

\textbf{One Lane per Road}: Suppose there is one lane per direction, and vehicles in each lane are only allowed to pass the intersection by going straight forward. Then the dimension of the state is $n=4$ and $\xvect := (x_1,\cdots,x_4)^T \in \Rbb^4$. We order the counts of the lanes in the intersection as a vector according to $(E,N,W,S)$. We further suppose that thhere is one designated traffic light assigned to each lane in each direction for a total of four lights, i.e. $\uvect := (u_1,\cdots,u_4)^T \in \{0,1\}^4$. However, certain constraints can be imposed on the four lights for basic efficiency, e.g. $E$ and $W$ traffic can pass simultaneously, and likewise for $N$ and $S$ traffic.

If we modeled the number of vehicles in each lane $x_\ell$, $\ell=1,\cdots, 4$, as a Markov chain then we take the transition probabilities to be given by
\begin{align*}
    p_{ij} = \begin{cases}
        p_Rp_0 + p_Gp_M &\text{ if } j = i\\
        p_Rp_{j-i} + p_Gp_{j-i+M} &\text{ if } j \in \{i+1, \cdots, i+M+1\}\\
        p_Rp_{j-i} &\text{ if } j \in \{i+M+2, i+x_{\text{max}}\}\\
        p_Gp_0 &\text{ if } j = \max\{0, i-M\}
    \end{cases}
\end{align*}
for all $i \in \Zbb^{\geq 0}$. Now suppose for concreteness, $M = 2$, $x_{\text{max}} = 5$, and $p_G = 0.6$, and suppose we are interested in the intersection snapshot $\xvect_0 := (5,1,5,1)^T$. Enumerating all possible one-timestep previous states $\xvect_0$ could have come from yields:
\begin{enumerate}
    \item $(y_1,x_1,y_2,x_2)^T$ where $y_1,y_2 \in \{2,3,4,5,6,7\}$ and $x_1,x_2 \in \{0,1\}$, which reach $\xvect_0$ in one step if the East and West lanes were given the green light at the same time some number of vehicles between five and zero entered the corresponding lanes, and if traffic in the North and South lanes were stopped at the same time one or no vehicles entered the correpsonding lanes.

    \item $(y_1,x_1,y_2,x_2)^T$ where $y_1,y_2 \in \{0,1,2,3,4,5\}$ and $x_1,x_2 \in \{0,1,2,3\}$, which reach $\xvect_0$ in one step if the East and West lanes were stopped at the same time some number of vehicles between five and zero entered the corresponding lanes, and if if traffic in the North and South lanes were given the green light at the same time three, two, one, or no vehicles entered the corresponding lanes.
\end{enumerate}

Then the expected time in between consecutive instances of $\xvect_0$ is given by
\begin{align}\label{eq:one_lane_first_time}
    \Ebb[T] := \left(\sum\limits_{\substack{x_1,x_2 \in \{0,1\}\\y_1,y_2 \in \{2,3,4,5,6,7\}}}\pi_{(y_1,x_1,y_2,x_2)}p_{y_{1,5}}p_{x_{1,1}}p_{y_{2,5}}p_{x_{2,1}} + \sum\limits_{\substack{x_1,x_2 \in \{0,1,2,3\}\\y_1,y_2 \in \{0,1,2,3,4,5\}}}\pi_{(y_1,x_1,y_2,x_2)}p_{y_{1,5}}p_{x_{1,1}}p_{y_{2,5}}p_{x_{2,1}}\right)^{-1}
\end{align}

Now we set $\xvect_0^{(1)} := \xvect_0$ and consider an alternative snapshot of the intersection $\xvect_0^{(2)} := (3,3,3,3)^T$. We are interested in the probability $P_1$ of observing $\xvect_0^{(1)}$ before $\xvect_0^{(2)}$ and the minimum time $T_{\text{min}}$ it takes to observe one of the two patterns. Denote $\Ebb[T_1] := \Ebb[T]$ from~\eqn{one_lane_first_time}. This is analogous to the coin tossing example of Section~\ref{subsec:pattern_occurrence}. 

Again enumerating all possible one-timestep previous states $\xvect_0^{(2)}$ could have come from yields:
\begin{enumerate}
    \item $(y_1,x_1,y_2,x_2)^T$ where $y_1,y_2 \in \{2,3,4,5\}$ and $x_1,x_2 \in \{0,1,2,3\}$, which reach $\xvect_0$ in one step if the East and West lanes were given the green light at the same time three, two, one, or no vehicles entered the corresponding lanes, and if traffic in the North and South lanes were stopped at the same time one or no vehicles entered the correpsonding lanes.

    \item $(y_1,x_1,y_2,x_2)^T$ where $y_1,y_2 \in \{0,1,2,3\}$ and $x_1,x_2 \in \{2,3,4,5\}$, which reach $\xvect_0$ in one step if the East and West lanes were stopped at the same time one or no vehicles entered the correpsonding lanes, and if if traffic in the North and South lanes were given the green light at the same time three, two, one, or no vehicles entered the corresponding lanes.
\end{enumerate}

Then the expected time in between consecutive instances of $\xvect_0^{(2)}$ is given by
\begin{align}\label{eq:one_lane_second_time}
    \Ebb[T] := \left(\sum\limits_{\substack{x_1,x_2 \in \{0,1,2,3\}\\y_1,y_2 \in \{2,3,4,5\}}}\pi_{(y_1,x_1,y_2,x_2)}p_{y_{1,3}}p_{x_{1,3}}p_{y_{2,3}}p_{x_{2,3}} + \sum\limits_{\substack{x_1,x_2 \in \{2,3,4,5\}\\y_1,y_2 \in \{0,1,2,3\}}}\pi_{(y_1,x_1,y_2,x_2)}p_{y_{1,3}}p_{x_{1,3}}p_{y_{2,3}}p_{x_{2,3}}\right)^{-1}
\end{align}

As in the coin tossing example of Section~\ref{subsec:pattern_occurrence}, we can set up the equations as follows. Note that there is no overlap between the two patterns, which implies $T_{1|2} = T_1$ and $T_{2|1} = T_2$.
\begin{align*}
    \Ebb[T_1] &= \Ebb[T_{\text{min}}] + (\Ebb[T_1] - \Ebb[T_{\text{min}}]) = \Ebb[T_{\text{min}}] + (1 - P_1)\Ebb[T_{1|2}] = \Ebb[T_{\text{min}}] + (1 - P_1)\Ebb[T_1]\\
    \Ebb[T_2] &= \Ebb[T_{\text{min}}] + (\Ebb[T_2] - \Ebb[T_{\text{min}}]) = \Ebb[T_{\text{min}}] + P_1\Ebb[T_{2|1}] = \Ebb[T_{\text{min}}] + P_1\Ebb[T_2]
\end{align*}

Hence, 
\begin{align*}
    P_1 = \frac{\Ebb[T_2]}{\Ebb[T_2] + \Ebb[T_1]}
\end{align*}
where $T_1$ is from~\eqn{one_lane_first_time} and $T_2$ is from~\eqn{one_lane_second_time}.

\textbf{One Lane per Road with Possible Jamming}: Suppose now there is the possibility of a traffic jam on the opposite side of the intersection, which might prevent vehicles from being able to pass the intersection even on a green light. Note that if the rate of cars clearing the intersection on a green light without jams is given by $M$, the rate of cars leaving from the opposite side of the intersection and clearing space should also be $M$. Furthermore, since jams are most likely caused by adjacent intersections, we let $p_G$ be the probability that the vehicles in the jammed lanes begin to move, and $p_R$ is the probability that they remain in place, preventing cars on the incident side of the intersection from being able to cross.

The sole difference between including and excluding lane jams is the transition probability matrix. Essentially, vehicles can only cross an intersection when both lights are green, and can be stopped at an intersection if the opposite lane is blocked with other cars. 
\begin{align*}
    p_{ij} = \begin{cases}
        (p_R + p_Gp_R)p_0 + p_G^2p_M &\text{ if } j = i\\
        (p_R + p_Gp_R)p_{j-i} + p_G^2p_{j-i+M} &\text{ if } j \in \{i+1, \cdots, i+M+1\}\\
        (p_R + p_Gp_R)p_{j-i} &\text{ if } j \in \{i+M+2, i+x_{\text{max}}\}\\
        p_G^2p_0 &\text{ if } j = \max\{0, i-M\}
    \end{cases}
\end{align*}
for all $i \in \Zbb^{\geq 0}$. We can then repeat the compuation shown in the previous scenario without lane jamming, with the appropriate Markov chain probabilities replaced.

\begin{figure}
    \begin{center}
        \includegraphics[width=0.4\columnwidth]{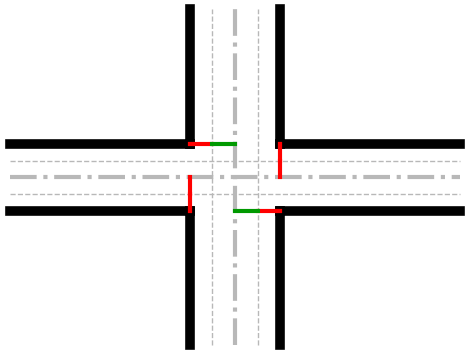}
    \end{center}
    \caption{A single four-way intersection with two lanes per direction. Red bars indicate that the corresponding lane is given the red light and stopped, while green bars indicate that vehicles in that lane are allowed to pass. In this particular configuration, the North and South left-turning lanes are given the green light.}
    \label{fig:4way}
\end{figure}

\textbf{Two Lanes per Road}: Now suppose there are two lanes per road: one for forward and right-turning traffic, and one for left-turning traffic. The dimension of the state is now $n=8$, $\xvect := (x_1,\cdots,x_8)^T \in \Rbb^8$, and we order of the components in the vector is `forward/right' first, `left' second with the directions ordered counterclockwise from East to South. We again assume that there is one designated traffic light per lane $\uvect := (u_1,\cdots,u_8)^T \in \{0,1\}^8$, but impose certain common-sense constraints as in the previous one-lane case, i.e., vehicle streams passing in opposite directions ($E$ and $W$, or $N$ and $S$) can pass simultaneously. We also assume that the roads of the intersection are wide enough so that left-turning vehicles from opposite directions can also pass at the same time. For a visualization of the intersection, we refer to~\fig{4way}. 

A vehicle can either choose to turn left, right, or go straight forward with the following probabilities:
\begin{enumerate}
    \item $p_L$ denotes the probability of vehicles which enter the left-turn lane among all the vehicles which enter the intersection from a neighboring intersection, and denote $p_{FR} := 1 - p_L$.

    \item $q_F$ denotes the probability that vehicles in the forward/right-turning lane of an intersection go forwards, and $q_R := 1 - q_F$ denotes the probability that a vehicle turns right.
\end{enumerate}

For each of the four directions, we let $p_{G,L}$ be the probability that vehicles in left-turning lane are given the green light, $p_{G,FR}$ be the probability that vehicles in forward and right-turning lane are given the green light, and $p_{R,L} := 1 - p_{G,L}$, $p_{R,FR} := 1 - p_{G,FR}$. Additionally, since there are different distances that need to be crossed depending on whether a vehicle is turning left, right, or going forward, constant speed vehicles still take different amounts of time to cross the intersection. Thus, we also distinguish the constants $M_L < x_{\text{max}}$ and $M_{FR} < x_{\text{max}}$. 

% Each vehicle is characterized by a tuple profile which keeps track of two objects: 1) its incident direction $\{ E, N, W, S\}$ and 2) its next intended action, either forward, right, or left. 

Then the Markov chain probabilities for all four forward and right-turning lanes are different from the probabilities for all four left-turning lanes. Namely, for the number of vehicles in each forward and right-turning lane $x_\ell$, $\ell=1,3,5,7$, the corresponding Markov chain transition probabilities are as follows:
\begin{align*}
    p_{ij} = \begin{cases}
        p_{R,FR}p_0 + p_{G,FR}p_{M,FR} &\text{ if } j = i\\
        p_{R,FR}p_{j-i} + p_{G,FR}p_{j-i+M_{FR}} &\text{ if } j \in \{i+1, \cdots, i+M_{FR}+1\}\\
        p_{R,FR}p_{j-i} &\text{ if } j \in \{i+M_{FR}+2, i+x_{\text{max}}\}\\
        p_{G,FR}p_0 &\text{ if } j = \max\{0, i-M_{FR}\}
    \end{cases}
\end{align*}
for all $i \in \Zbb^{\geq 0}$ and for the number of vehicles in each left-turning lane $x_\ell$, $\ell=2,4,6,8$, the corresponding transition probabilities are:
\begin{align*}
    p_{ij} = \begin{cases}
        p_{R,L}p_0 + p_{G,L}p_{M,L} &\text{ if } j = i\\
        p_{R,L}p_{j-i} + p_{G,L}p_{j-i+M_{L}} &\text{ if } j \in \{i+1, \cdots, i+M_{L}+1\}\\
        p_{R,L}p_{j-i} &\text{ if } j \in \{i+M_{L}+2, i+x_{\text{max}}\}\\
        p_{G,L}p_0 &\text{ if } j = \max\{0, i-M_{L}\}
    \end{cases}
\end{align*}

\section{Conclusion}\label{sec:conclusion}
In this paper, we extended the study of stochastic control theory to include the class of Poisson shot noise by designing a two-part hierarchical control policy for systems of the form~\eqn{gen_shot_sde} and~\eqn{gen_levy_sde}.
% In particular, we considered two problems of interest for stochastic systems of these types: 1) incremental stability analysis and 2) controller synthesis. The incremental stability problem is discussed in more detail in the first part of this paper \blue{(cite first paper)}, and essentially characterizes a nonlinear incremental stability framework for systems of the form~\eqn{gen_shot_sde} and~\eqn{gen_levy_sde} using contraction theory. Results demonstrate that the expected mean-squared difference between trajectories of the system with different initial conditions and noise process sample paths still converge exponentially towards within a bounded steady-state error ball of each other.
% We leveraged the results obtained from the stability criteria of the first part of this paper~\cite{han21tac} to , in address of the second controller synthesis problem. 
The first part of the control policy invokes renewal theory to construct a learning component which recognizes previously-occurred states and their corresponding optimal control responses. Namely, we utilize the ``pattern-occurrence'' problem to determine the expected time between consecutive instances of the same state. Even incorporating this simple learning component into the system has the potential to save time and computational energy because there is no need to devote resources towards recomputing a control action for a state that has been observed before. The second part of the control policy invokes an impulse control approach which computes the actual control action which is optimal for each of the specific states kept track of by the first learning component part. 

We then presented two simplified specific applications for control design: 1) a 1D fault-tolerance system and 2) a control problem for moderating vehicle traffic at an intersection. For the 1D fault-tolerance problem, we explicitly computed steady-state error bound $\kappa_s(\beta_s,t)$ and showed that larger jump norm bounds $\eta$ and 2) shorter interarrival times between jumps correspond to a larger error ball. We then used renewal theory to address the converse problem of determining the largest possible values of the shot noise characteristics, i.e. intensity and maximum absolute jump norm, such that the steady state error bound remains bounded by a desired tolerance. In this context, the renewals referred to the event where the trajectory exceeds the stability bound by an amount which makes the system unstable. The impulse control method is then applied at these points to navigate the trajectory back within the desired error bound. 
%irrecovereable

For the vehicle traffic control problem, we used renewal theory to compute the expected time between consecutive renewals, which referred to certain snapshots of the intersection. The reason for saving these snapshots and their corresponding optimal control actions is to save resources. One might observe 10 cars in the East lane and five cars in the North lane, and the optimal control action might be to let the first five East cars pass, then the five North cars, and finally the last five East cars. If the system observes the exact same scenario in a future time, we simply recycle the same optimal control action used previously. We showed that renewal theory is also useful for choosing which states are worth saving in memory. For snapshots which do not have a high probability of occurring (e.g., ten vehicles occupying all four directions of the intersections simultaneouly), there is not much computational benefit to saving the corresponding optimal control action, especially when the system has limited storage. The performance objective for the impulse control method refers to the level of congestion in the intersection, and we desire to compute an action of traffic light sequences which minimizes the level of congestion as quickly as possible. Unlike the 1D reference-tracking problem, a necessary addition to the standard impulse control method is a time-varying performance criteria which assigns varying priorities to the lanes depending on the number of vehicles that are present at each lane.

% We emphasize that the benefits of our work are two-fold: 1) the phenomenon of jumps in noise is understudied in the controls community compared to Gaussian white noise despite being equally prevalent, and 2) it segways into a methodological design process for stochastic controllers and observers that counter noise processes which include instantaneous, large-magnitude jumps, without necessarily resorting to computationally intensive model-free techniques simply because the noise process is non-Gaussian. 

We argue that the simplified examples investigated in this paper are sufficient enough to develop the theoretical framework for the controller synthesis procedure. To make the controller more applicable to more complex real-life instances, certain factors also need to be considered. One extension is towards the direction of continuous-valued jumps, which may be addressed by first discretizing the space of jumps:
\begin{align*}
    \Scal := \{ \xi(t,\xvect,\zvect) | \norm{\xi(t,\xvect,\zvect)} \leq \eta\} \ \Longrightarrow \ \Scal_i := \{ \xi(t,\xvect,\zvect) | \norm{\xi(t,\xvect,\zvect)} \leq \eta_i\} 
\end{align*} 
for chosen threshold values $0 < \eta_1 < \eta_2 < \cdots \leq \eta_k := \eta$, where $\eta$ is the upper bound on the norm of the jumps, and some $k\in\Nbb$. Then any two jumps $\xi_a(t,\xvect,\zvect), \xi_b(t,\xvect,\zvect)$ which belong in the same set $\Scal_i$ can be controlled using the same law. Another possible extension to the pattern occurrence problem is when we care about what specific order a certain number of patterns $A_1, \cdots, A_M$ occurs in. For example, for the case of $M=3$, define $T^{*}$ to be the expected time until we see $A_3$ after we have already observed occurrences of $A_1, A_2$ in any order. Then, with the notation $P_i$ being the probability of observing pattern $A_i$ first among all $M$ patterns, we have
\begin{align*}
	\Ebb[T^{*}] = P_{A_1}(\Ebb[T_{A_1}] + \Ebb[T_{A_2|A_1}]) + P_{A_2}(\Ebb[T_{A_2}] + \Ebb[T_{A_1|A_2}])
\end{align*}
The interest for looking into these types of pattern arrangements arises from applications in fault tolerance such as the turbulence of ocean waves affecting a drifting boat, or the collapse of the Tacoma Narrows Bridge in 1940. While these systems may have been designed to withstand strong perturbations, it may fail in events resulting from a steady accumulation of consecutive smaller perturbations. An extension for the modulation control component is the replacement of the impulse controller with an alternative controller, especially in applications where a sharp decay in the state is not physically feasible. The hierarchical structure of the two-part controller remains the same regardless: the learning component facilitates the recognition of destabilizing patterns, and the modulation controller is applied only then. One could also derive first-order conditions to impose additional limits on the extent of being able to control the system using impulsive jumps of large magnitude.
% Not only a bound on position deviation away from desired trajectory
% \begin{itemize}
    % \item \textbf{Practicality of the impulse controller}: impulse isn’t actually impulse, but hill. Closest practical working example is bang-bang control in spacecraft fault tolerance. The novel contribution of the paper is the two-part construction of the controller, and proposing the learning component in order to reduce redundancy in computation.
    % \item epsilon-stability probability bounds for stability. No longer a strong incremental stability condition.
% \end{itemize}

% % \violet{inefficiency of neural networks...} 

%%%%%%%%%%%%%%%%%%%%%%%%%%%%%%%%%%%%%%%%%%%%%%%%%%%%%%%%%%%%%%%%%%%%%%%
%%%%%%%%%%%%%%%%%%%%%%%%%%%%%%%%%%%%%%%%%%%%%%%%%%%%%%%%%%%%%%%%%%%%%%%

\section*{Acknowledgments}
The authors would like to thank John C. Doyle for the discussion of ideas which provided motivation for this work.

%%%%%%%%%%%%%%%%%%%%%%%%%%%%%%%%%%%%%%%%%%%%%%%%%%%
%%%%%%%%%%%%%%%%%%%%%%%%%%%%%%%%%%%%%%%%%%%%%%%%%%%
%%%%%%%%%%%%%%%%%%%%%%%%%%%%%%%%%%%%%%%%%%%%%%%%%%%
\newpage
%%%%%%%%% BIBLIOGRAPHY %%%%%%%%%%%%%%%%%%%%%%%%%%%%%%%%%
\bibliographystyle{apa}
\bibliography{%
\bibi%
}
\end{document}